\documentclass[useAMS,usenatbib,usenatNishiNishibib]{mn2e}

\usepackage{graphicx}

\newcommand{\dpr}{$^{\prime\prime}$}
\newcommand{\pr}{$^{\prime}$}
\newcommand{\hi}{H~{\sc i} }
\newcommand{\hii}{H~{\sc ii} }

\newcommand{\dg}{$^{\circ}$}

\newcommand{\nii}{[N~{\sc ii}] }
\newcommand{\ha}{$H\alpha$ }
\newcommand{\hb}{$H\beta$ }

\title[Tidal tails of interacting galaxies]{Star-forming regions and the metallicity gradients in the tidal tails: The case of NGC 92\thanks{Based on observations obtained at the Gemini Observatory, which is operated
by the Association of Universities for Research in Astronomy, Inc.,
under a cooperative agreement with the NSF on behalf of the Gemini partnership:
the National Science Foundation (United States), the Science and
Technology Facilities Council (United Kingdom), the National Research
Council (Canada), CONICYT (Chile), the Australian Research Council
(Australia), Minist\'erio da Ci\^encia e Tecnologia (Brazil) and Ministerio de
Ciencia, Tecnolog\'ia e Innovaci\'on Productiva (Argentina) -- Observing runs:
GS-2006B-Q-79 and GS-2011B-Q-36.}}

\author[S. Torres-Flores et al.]
{
\parbox[t]{\textwidth} {S. Torres-Flores$^{1}$\thanks{E-mail: storres@dfuls.cl}, S. Scarano Jr$^{2,3,4}$, C. Mendes de Oliveira$^{3}$, D. F. de Mello$^{5}$, P. Amram$^{6}$ \& H. Plana$^{7}$}
\vspace*{6pt}\\
$^1$Departamento de F\'isica, Universidad de La Serena, Av. Cisternas 1200 Norte, La Serena, Chile \\
$^2$Instituto de Astronomia, Geof\'{\i}sica e Ci\^encias Atmosf\'ericas da Universidade de S\~ao Paulo,\\ Cidade Universit\'aria, CEP: 05508-900, S\~ao Paulo, SP, Brazil \\
$^3$Southern Astrophysical Research Telescope (SOAR), Casilla 603, La Serena, Chile\\
$^4$Departamento de F\'isica - CCET, Universidade Federal de Sergipe, Rod. Marechal Rondon s/n, 49.100-000, \\Jardim Rosa Elze, S\~ao Cristov\~ao, SE, Brazil\\
$^5$Catholic University of America, Washington, DC 20064, USA\\
$^6$Laboratoire d'Astrophysique de Marseille, Aix Marseille Universit\'e, CNRS, 13388, Marseille, France\\
$^7$Laborat\'orio de Astrof\'isica Te\'orica e Observacional, Universidade Estadual de Santa Cruz, Ilh\'eus, Brazil
}

\begin{document}
\maketitle

\begin{abstract}

We present new Gemini/GMOS spectroscopic and archival imaging data of the interacting galaxy NGC 92, which is part of a compact group and displays an extended tidal tail. We have studied the physical properties of 20 star-forming complexes in this system. We found that the star-forming regions located in the tidal tail of NGC 92 have ages younger than $\sim$8 Myr, which suggests that these objects were formed \textit{in situ}. The spectroscopic data reveals that these regions have slightly sub-solar metallicities, suggesting that they were formed from pre-enriched material. Using the oxygen abundances derived for each system, we found that the extended tidal tail of NGC 92 has a flat metallicity distribution. Although this scenario is consistent with N-body simulations of interacting systems, where there is gas mixing triggered by the interaction, archival H$\alpha$ Fabry-Perot data cubes of NGC 92 have not shown a velocity gradient along the tail of this galaxy, which under certain assumptions could be interpreted as a lack of gas flow in the tail. Our results suggest that a fraction of the enriched gas that was originally located in the center of the galaxy was expelled into the tidal tail when the interacting process that formed the tail happened. However, we can not exclude the scenario in which the star formation in the tail has increased its original oxygen abundance.

\end{abstract}


\section{Introduction}

Interacting systems have been used as an useful tool to understand the formation and evolution of galaxies. Several phenomena that take place in interacting galaxies located in the nearby Universe can be extrapolated to the high-z Universe. For example, the chemical evolution of distant galaxies can be studied by using local interacting objects. \cite{Cresci10} and \cite{Queyrel12} found that high-z galaxies display flat or inverted metallicity gradients, a phenomenon that has been found in local interacting systems \citep{Werk11}. In the same context, \cite{Kewley10} studied a sample of local close galaxy pairs. These authors found that pairs show flatter metallicity gradients than isolated galaxies. Some authors have used stellar cluster to trace the chemical abundance of interacting galaxies. For example, \cite{Schweizer98} studied the spectra of eight clusters in the merger remnant NGC 7252, suggesting that some of them were formed at the beginning of the merging processes. \cite{Chien07} studied a sample of 12 star clusters located in the northern tidal tail of the system NGC 4676. Using the R$_{23}$ method, they found a flat metallicity distribution for this tail, suggesting an efficient gas mixing process. However, the R$_{23}$ metric should be used with caution in moderate signal-to-noise spectra, as it can be misleading. Recently, \cite{Trancho07a,Trancho07b}, \cite{Bastian09} and \cite{Trancho12} have studied the population of young star clusters in merging/interacting galaxies. Using Gemini/GMOS imaging and spectroscopic data, these authors used clusters to trace the interaction history and evolution of the parent galaxies. These authors have found several young clusters in tidal tails, e. g. in NGC 3256 \citep{Trancho07a,Trancho07b}. Despite the fact that tidal tails are typical features of interacting/merging galaxies, there is a limited number of studies regarding the metallicity gradients of these structures. At this point, several questions arise: what should be the trend of the metallicity gradients along tidal tails? How does the metal transport in tidal tails work? In order to address these questions, we have started an imaging and spectroscopic program devoted to determine the physical properties of star-forming regions in the tidal tails of interacting systems, how the metal mixing works in these structures and how it correlates with the metallicity of the disk of the main interacting galaxies. The first system in our analysis is galaxy NGC 92, a member of a compact group first catalogued by \cite{Rose77} as Rose 34, later called Robert's quartet.

Robert's quartet corresponds to a group of late-type galaxies, which is located at a distance of 44.7 Mpc \citep{Mould00}. The members of this system are NGC 87 (Irr), NGC 88 (SAab pec), NGC 89 (SB0/a pec) and NGC 92 (SAa pec) and all of them present clear signatures of interactions (see \cite{Presotto10}). In fact, in this compact group, the HI gas forms a common envelope around the galaxies {and a bridge} between NGC 92 and NGC 88 \citep{Pompei07}, which suggests that galaxy-galaxy interactions have taken place in this system. In the case of NGC 92, the main galaxy of this system, it is classified as a LINER \citep{Coziol00} and it has an extended optical and gaseous tidal tail. Torres-Flores et al. (2009) analyzed the far and near ultraviolet \textit{GALEX} images of NGC 92 and found several UV sources in this tidal tail. Given that the UV light is closely related to young stellar populations, this tidal tail is a good laboratory to study the physical properties of young star-forming regions located in tidal features, specially chemical abundances. These abundances can be used to study metallicity gradients along these structures. In this sense, this analysis can help us to determine if tidal tails play an important role in the chemical enrichment of interacting systems. In order to study the physical properties of these young stellar associations (e. g. ages, internal extinctions, line ratios, oxygen abundances) we have obtained new Gemini GMOS spectroscopic data for NGC 92 and we retrieved Gemini GMOS archival images of this system. In this paper, we assumed 43.2 Mpc as the distance to NGC 92, considering the Hubble flow, \mbox{H$_0$=73 km s$^{-1}$ Mpc$^{-1}$} and the systemic velocity given by \cite{RC3}.

This paper is organized as follows. In Sections 2 and 3 we present the data and data analysis. In Section 4 we present the results. In Section 5 we discuss our results and in Section 6 we present our main conclusions.

\section{Observations}

\subsection{Data and data reduction}

\subsubsection{GMOS imaging}

NGC 92 was observed with the Gemini MultiObject Spectrograph (GMOS) at the Gemini South telescope in 2006 under the science programme GS-2006B-Q-79 (PI: W. Harris). This object was observed in the \textit{u\pr}, \textit{g\pr} and \textit{r\pr}-band filters. A set of dithered images was taken in each filter in order to eliminate the detector gaps and cosmic rays. Two of the four images observed in \textit{g\pr} filter were observed in different nights, but under similar observational conditions. This also happened with two of the seven images observed with the \textit{u\pr}-band filter. All images observed in the \textit{r\pr} filter are from the same night. There are also \textit{i\pr}-band observations of this system, but none of the calibration images  observed at that time or the images from the Gemini archive were suitable to eliminate the fringing pattern. All images were observed with an instrumental position angle of 141\dg and with a 2$\times$2 binning mode. Details on each observation can be seen in Table \ref{tbl-1}.

Gemini GMOS images were reduced by using the IRAF\footnote{IRAF is distributed by the National Optical Astronomy Observatories, which are operated by the Association of Universities for Research in Astronomy, Inc., under cooperative agreement with the National Science Foundation. See http://iraf.noao.edu} Gemini Data Reduction software. Individual images were corrected by overscan level, differences in gain of each detector, bias and flat-field using GBIAS, GIFLAT and GIREDUCE, in a similar way as described by \cite{Sca2008a}. For each filter, the multiple dithered frames were mosaiced and co-added with the IMCOADD task. Zero-point photometric calibrations were taken from the Gemini Baseline Calibration based on \cite{Ryder+06}.

\begin{table}
\centering
\begin{minipage}[t]{0.5\textwidth}
\caption{Gemini/GMOS imaging details}
\begin{tabular}{lrrr}
\hline
\multicolumn{ 1}{l}{{\bf Parameter}}         &   \multicolumn{ 1}{c}{{\bf u\pr}} &   \multicolumn{ 1}{c}{{\bf g\pr}} &   \multicolumn{ 1}{c}{{\bf r\pr}} \\
\hline
UT Date\footnote{Universal date and time when the observation started.}           & 23/09/2006   & 23/09/2006   & 23/09/2006   \\
UT Time$^{\textit{a}}$           & 03:28:26.6   & 02:12:31.1   & 04:01:37.6   \\
UT Date$^{\textit{a}}$            & 29/10/2006   & 29/10/2006   &      -       \\
UT Time$^{\textit{a}}$            & 05:00:02.0   & 05:06:33.5   &      -       \\
UT Date$^{\textit{a}}$            & 24/11/2006   & 24/11/2006   &      -       \\
UT Time$^{\textit{a}}$            & 01:17:01.0   & 01:23:32.5   &      -       \\
t$_{exp}$ [s]\footnote{Exposure time.}      &        330   &        321   &        320   \\
n$_{exp}$\footnote{Number of exposures for the dithering pattern.}          &          7   &          4   &          4   \\
Airmass            &       1.10   &       1.14   &       1.06   \\
Seeing [\dpr]      &       0.98   &       0.72   &       0.73   \\
\hline
\vspace{-0.8cm}
\label{tbl-1}
\end{tabular}
\end{minipage}
\end{table}

\subsubsection{GMOS spectroscopy}

\begin{figure*}
\includegraphics[width=0.9\textwidth]{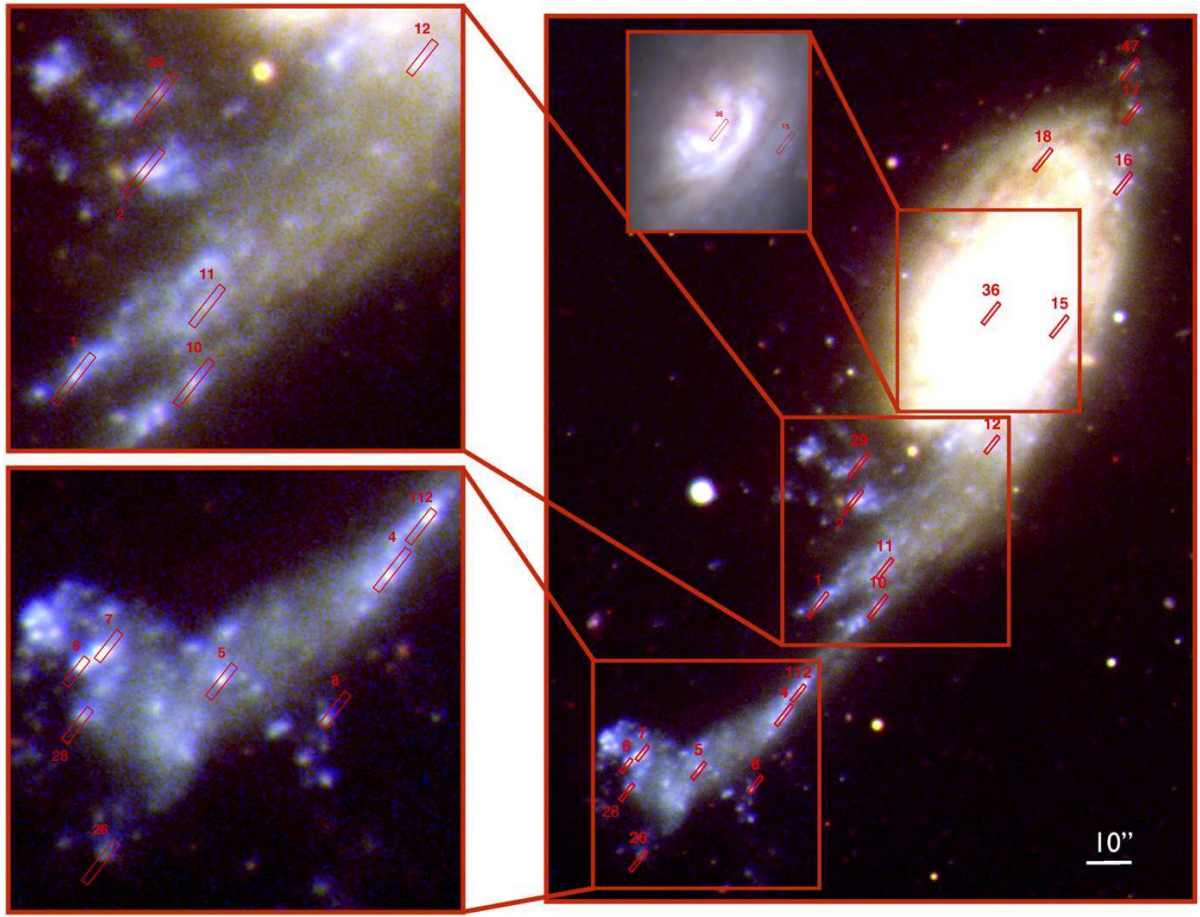}
\caption{\textit{u\pr}, \textit{g\pr} and \textit{r\pr}-band composite Gemini/GMOS image of NGC 92, where blue, green and red colours are associated with each filter, respectively. Red circles indicated the regions for which we have obtained spectroscopic data.}
\label{ngc92_rgb_regions_phot_paper_2}
\end{figure*}

Spectroscopic observations were carried out on 2011 September 30, under the science programme GS-2011B-Q-36 (PI: S. Torres-Flores) at the Gemini South telescope in queue mode using the GMOS spectrograph. Given that u, g and r-band images of NGC 92 were already available from the archive, we used these images to select the bluest star-forming regions in the tidal tails of NGC 92. In Fig. \ref{ngc92_rgb_regions_phot_paper_2} we show the \textit{u\pr}, \textit{g\pr} and \textit{r\pr}-band Gemini/GMOS image of NGC 92, where the red circles indicate the regions for which we have obtained spectroscopic data. We compared the spatial position of these optical knots with the \textit{GALEX} images of NGC 92 and we found a good agreement between the position of the optical knots and the peaks in the UV emission. We then obtained a catalogue with the position of the star-forming regions candidates. With this information in hand, we designed the GMOS mask to obtain the spectroscopy of the selected targets. Since our sources are mostly pointlike we adopted slit widths of 1 arcsec to maximize the incoming light and minimize the background contamination given the image quality requirements we had (better than 0.9 arcsec). The size of the slits were set individually with the purpose of improving the number of observed \hii regions and the local sky sampling. All slits were oriented according to the instrumental position angle of 141\dg.

The observational strategy was to perform exposures of 1200s in four different central wavelengths, ranging from 6750-6850 \AA, in intervals of 50\AA, using the R400$\_$G5325 grism. This was done in order to cover the gaps of the GMOS CCDs. Depending on the position of the slit on the mask, different spectral coverages were achieved (from 4015-7775 {\AA} to 5415-9790 {\AA}), and the resolution was 4.5 {\AA} at H$\alpha$ (R$\sim$1500). The rms in the wavelength calibration was 0.38 {\AA}. These values were optimized to obtain \nii and \ha with the maximum signal in all spectra, in order to estimate the oxygen abundances using the [N~{\sc ii}]/$H\alpha$ calibrator by \cite{PP04} and \cite{Stasinska06}. To avoid any effect caused by instrumental flexures, CuAr arc lamps and flat-field frames were observed after spectroscopic exposures, while the telescope was still following the source. Details on the spectroscopic observations can be found in Table \ref{tbl-3}.

\begin{table}
\centering
\begin{minipage}[t]{\columnwidth}
\scriptsize
 \caption{Gemini/GMOS spectroscopic instrumental and observational setup}
 \label{tbl-3}
\begin{tabular}{rrrrr}
\hline
{\bf Parameter} & {\bf Exp. 1} & {\bf Exp. 2} & {\bf Exp. 3} & {\bf Exp. 4} \\
\hline
$\lambda_c$ [\AA] \footnote{Central wavelength of the dispersed light.}   &       6700 &       6750 &       6800 &       6850 \\
UT Date\footnote{Universal date and time when the observation started.}             & 30/09/2011 & 30/09/2011 & 30/09/2011 & 30/09/2011 \\
UT Time$^{\textit{a}}$             &   05:17:00 &   04:31:00 &   04:02:05 &   04:52:34 \\
Range [\AA]\footnote{Spectral range coverage relative to the center of the GMOS field.}                            &  4600-8800 &  4650-8850 &  4700-8900 &  4750-8950 \\
Airmass\footnote{Mean airmass during the observation.}                     &       1.27 &       1.16 &       1.12 &        1.20 \\
t$_{exp}$ [s]\footnote{Exposure time.}                                     &       1200 &       1200 &       1200 &       1200 \\
Calib. Star\footnote{Spectrophotometric star for flux calibration.}        &    LTT9239 &    LTT9239 &    LTT9239 &    LTT9239 \\
\hline
\vspace{-0.8cm}
\end{tabular} 
\end{minipage}
\end{table}

Typical seeing during the observations was better than 0.9 arcsec under dark and photometric conditions. In queue mode it is not possible to optimize MOS spectroscopy to the parallactic angle. Fortunately, in our case, the observations were conducted in a mean angle lower the 30{\dg} relative to this angle. Gemini Observatory webpage shows a detailed discussion on atmospheric dispersion effects as a function of the angle between the slit and the parallactic angle\footnote{http://www.gemini.edu/?q=node/11212}. Using the IDL code DIFF$_{-}$ATM$_{-}$REFR by Enrico Marchetti, we simulated the amount of light lost due to atmospheric refraction. Taking into account the slit width in our worst observational conditions (see section 2.1.2) light losses were about 5\%.

The data reduction was applied using the Gemini Data Reduction package. Each frame was corrected by overscan, bias and flat-field according to the observed central wavelengths using GBIAS, GSFLAT and GSREDUCE. The most severe cosmic ray events were removed using LACOS$_{-}$SPEC script by van Dokkum (2001). Individual 2D spectra, associated to each slit in each central wavelength, was cut from the MOS field using GSCUT task and combined to remove the detector gaps and the remaining cosmic rays using GEMCOMB task. Bidimensional solutions for wavelength calibration were found for all slits using GSWAVELENGTH and then transposed to the science spectra using GSTRANSFORM task. Continuum and sky lines were removed from these spectra with GSSKYSUB task. The spectral dispersion was traced and extracted. When the continuum was not detectable, tracing was performed using the emission-line positions. In this case, a two-sigma cut level from the local background of the strongest line in the spectrum was used to constrain the aperture for task TRACE. Gap and skyline artefacts were removed by interpolating bad data ``case-by-case'' using IRAF task SPLOT. Following the similar procedures, we reduced the spectra of the standard star LTT9239 to determine the sensitivity function in our observations and to flux calibrate all spectra. The final spectra for the regions belonging to NGC 92 are shown in Figs. \ref{plot_spectra1}, \ref{plot_spectra2} and \ref{plot_spectra3} we show these spectra. 

\begin{center}
\begin{table}
\begin{minipage}[t]{0.5\textwidth}
\caption{Position and velocities of the regions with GMOS data}
\begin{tabular}{cccccccccc}
\hline
ID & R.A. & Dec. & Velocity \\
   & J2000 & J2000 & km s$^{-1}$  \\

\hline
   TF1  &  00:21:35.13  &   -48:38:25.7   &      3268.4$\pm$ 52.3    \\ 
   TF2  &  00:21:34.45  &   -48:38:06.0   &      3233.9$\pm$ 29.4    \\
   TF4  &  00:21:35.89  &   -48:38:48.2   &      3228.3$\pm$ 27.8    \\
   TF5  &  00:21:37.54  &   -48:38:57.9   &      3280.4$\pm$ 36.3     \\
   TF6  &  00:21:38.92  &   -48:38:56.4   &      3254.1$\pm$ 37.1     \\
   TF7  &  00:21:38.72  &   -48:38:55.2   &      3292.4$\pm$ 48.3     \\
   TF8  &  00:21:36.52  &   -48:39:02.0   &      3227.2$\pm$ 46.1     \\ 
  TF10  &  00:21:33.99  &   -48:38:26.6   &      3307.5$\pm$ 31.6     \\
  TF11  &  00:21:33.87  &   -48:38:19.2   &      3273.5$\pm$ 83.3     \\
  TF12  &  00:21:31.76  &   -48:37:55.8   &      3274.2$\pm$ 59.6     \\
  TF15  &  00:21:30.41  &   -48:37:32.5   &      3438.4$\pm$ 60.7     \\
  TF16  &  00:21:29.17  &   -48:37:05.8   &      3537.2$\pm$ 24.1     \\
  TF17  &  00:21:28.81  &   -48:36:50.5   &      3550.9$\pm$ 20.0     \\
  TF18  &  00:21:30.57  &   -48:36:58.9   &      3571.3$\pm$ 68.4    \\ 
  TF26  &  00:21:38.70  &   -48:39:15.0   &      3270.7$\pm$ 56.1     \\
  TF28  &  00:21:38.90  &   -48:39:01.5   &      3247.1$\pm$ 47.4     \\
  TF29  &  00:21:34.31  &   -48:37:58.4   &      3247.6$\pm$ 44.8     \\
  TF36  &  00:21:31.64  &   -48:37:29.0   &      3359.7$\pm$ 51.1     \\
  TF47  &  00:21:28.87  &   -48:36:41.9   &      3472.1$\pm$ 36.1     \\
  TF112  &  00:21:35.51  &   -48:38:42.4   &      3266.5$\pm$ 22.6     \\
\hline
\vspace{-0.8cm}
\label{table3}
\end{tabular}
\end{minipage}
\end{table}
\end{center}

\subsubsection{H$\alpha$ Fabry-Perot data}

We have complemented the Gemini/GMOS observation of NGC 92 with archival $H\alpha$ Fabry-Perot observations for this system, which were already published in \cite{TorresFlores09}. These observations were carried out with the Fabry-Perot instrument CIGALE, at the European Southern Observatory 3.6 m telescope (ESO) in September 2000, with a spectral resolution of \mbox{$\sim$12 km s$^{-1}$} and a pixel scale of 0.405 \mbox{arcsec pixel$^{-1}$}, under a seeing of $\sim$1.0 arcsec. Details about the data reduction can be found in \cite{TorresFlores09}.

\section{Analysis}

\subsection{Photometry}

The main focus of this study is the spectroscopic analysis of star-forming regions located in tidal tails and in the main body of an interacting galaxy. However, the good image quality of the Gemini/GMOS data allowed us to also perform a photometric analysis of the regions away from the galaxies in particular for color determination. Given that our photometric sources correspond to the same sources for which we have spectroscopic data, we are able to compare the physical properties derived from the photometry and the spectroscopy (e. g. ages and internal extinctions). We note that in the photometric analysis we have added three sources that were not included in the GMOS mask. 

The \textit{u\pr}, \textit{g\pr} and \textit{r\pr}-band photometry of the star-forming regions was performed by using the task PHOT in IRAF. To do this, we used a fixed aperture of 0.5\arcsec radius, centered on the centroid of the emission on each band. This aperture was selected to match the slit width of the spectroscopic observations (1.0\arcsec). The sky was subtracted by using a sky annulus, where the parameters ANNULUS (inner radius of sky annulus) and DANNULUS (width of sky annulus) were set to 16 and 3 pixels (2.3\arcsec and 0.4\arcsec), respectively, and the sky fitting algorithm was the mode of the values inside the annulus, in a similar way as done by \citep{TorresFlores12} for NGC 2782. The \textit{u\pr}, \textit{g\pr} and \textit{r\pr}-band magnitudes were corrected by Galactic extinction using $A_{u}= E(B-V) \times 4.79$, $A_{g}= E(B-V) \times 3.81$ and $A_{r}= E(B-V) \times 2.74$, respectively, where the extinction coefficients were taken from \citep{Savage79} and the E(B-V) values were taken from \citep{Schlegel98}.

\subsection{Photometric ages and internal extinctions}

We have derived the photometric ages and the internal extinction of the star-forming regions detected in the tidal tail of NGC 92 by comparing the (\textit{u\pr}-\textit{g\pr}) and (\textit{g\pr}-\textit{r\pr}) observed colours with respect to the single stellar population models obtained from the Starburst99 database (hereafter, SB99) \citep{Leitherer99}. We generated SB99 models, for an instantaneous burst, Salpeter initial mass function (IMF; 0.1-100 M$_{\odot}$), solar metallicity, from 1 Myr to 1 Gyr. We note that these models does not include the explicitly the nebular emission associated with the H$\alpha$ line (see \S 4.2). Finally, these models were tuned for the \textit{u\pr}, \textit{g\pr} and \textit{r\pr}-band filters, which allow us to obtain the modeled (\textit{u\pr}-\textit{g\pr}) and (\textit{g\pr}-\textit{r\pr}) colours. We note that the ages derived from observed broad-band colors can be affected by an incomplete sampling of the IMF, which is not the case for the SB99 models, which assume a fully sampled IMF. Then, the comparison of observed and modeled colours can produce uncertainties in the age estimation of low-mass systems \citep[see][]{Cervino03,Hancock08,daSilva12}. Also, it is well known that optical broad band colours are degenerated for ages larger than 10 Myrs \citep[e. g.][]{Trancho12}. A way to break this degeneracy is to include ultraviolet photometry, however, our star-forming complexes are not resolved in the \textit{GALEX} data of NGC 92. Given that the regions detected in the tidal tail of NGC 92 are H$\alpha$ emitting line regions, we can assume that they are younger than 10 Myrs (for an instantaneous burst). For this reason, we have compared the observed colours with models that range from 1 to 8 Myrs. This fact allows us to avoid the age/extinction degeneracy for these systems. In order to determine the internal extinction of the star-forming regions (i. e. the color excess of each region), we have used the starburst extinction law defined by \cite{Calzetti94}, where we used $A_{u}= E(B-V) \times 6.08$, $A_{g}= E(B-V) \times 4.73$ and $A_{r}= E(B-V) \times 3.54$, respectively. Finally, ages and internal extinctions were estimated simultaneously by minimizing the $\chi^{2}$ between the observed and modeled (\textit{u\pr}-\textit{g\pr}) and (\textit{g\pr}-\textit{r\pr}) colours.

\begin{figure*}
\includegraphics[width=0.90\textwidth]{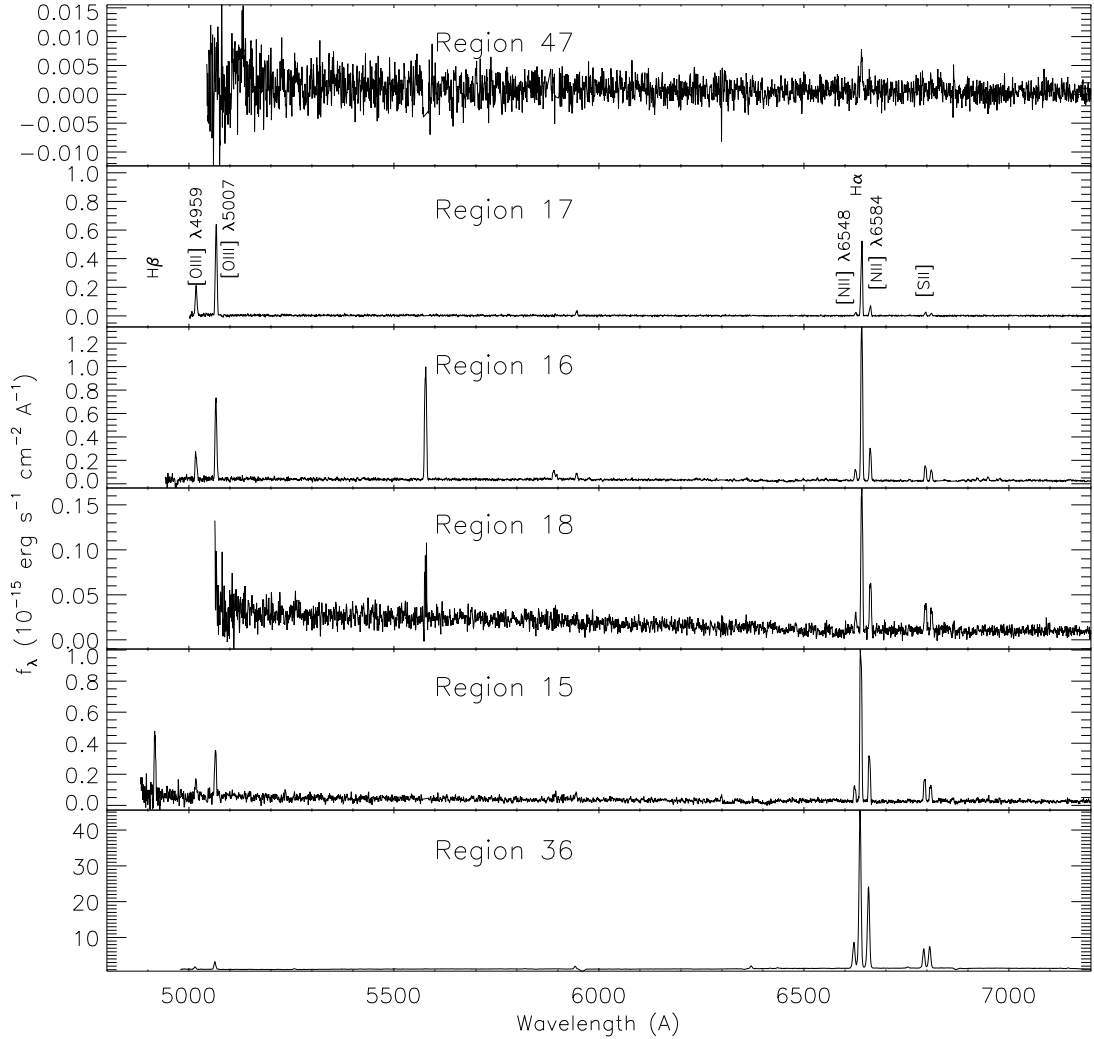}
\caption{Gemini/GMOS spectra of the star-forming regions located in the main body of NGC 92. The spectra are plotted starting from the northern to the southern regions (see Fig. \ref{ngc92_rgb_regions_phot_paper_2}, where the North is at the top and East is to left). The spectra of regions 47, 17, 16, 18 and 36 did not cover the H$\beta$ emission line. Some interpolation feature may be seen when gap and skyline artefacts were removed with the procedure presented in \S 2.1.2.}
\label{plot_spectra1}
\end{figure*}

\begin{figure*}
\includegraphics[width=0.90\textwidth]{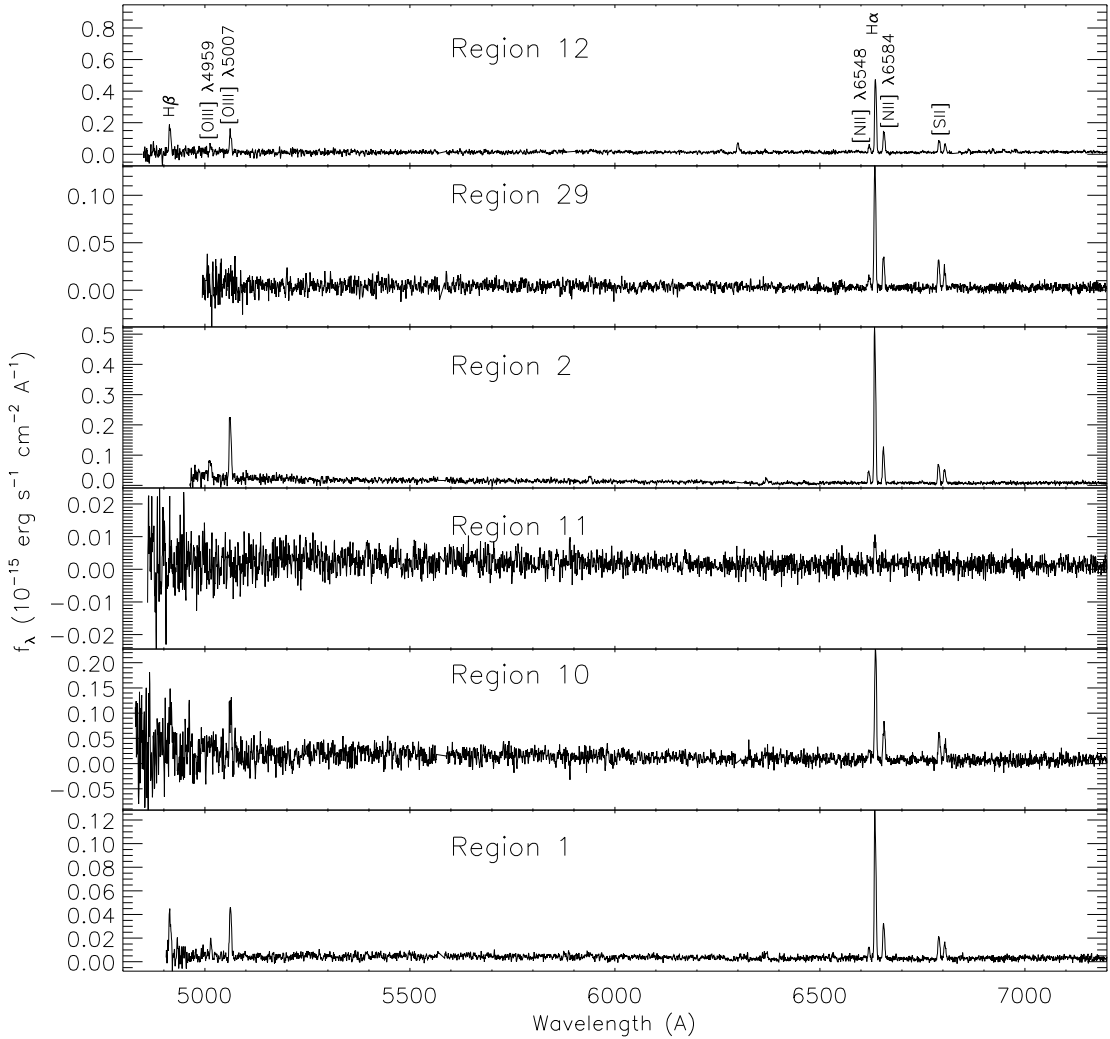}
\caption{Gemini/GMOS spectra of the star-forming regions located across the tidal tail of NGC 92. In this figure we show the spectra of regions that are shown in the central close-up of Fig. \ref{ngc92_rgb_regions_phot_paper_2}, starting from the northern to the southern regions. The spectra of regions 29 and 2 did not cover the H$\beta$ emission line. Some interpolation feature may be seen when gap and skyline artefacts were removed with the procedure presented in \S 2.1.2.}
\label{plot_spectra2}
\end{figure*}

\begin{figure*}
\includegraphics[width=0.90\textwidth]{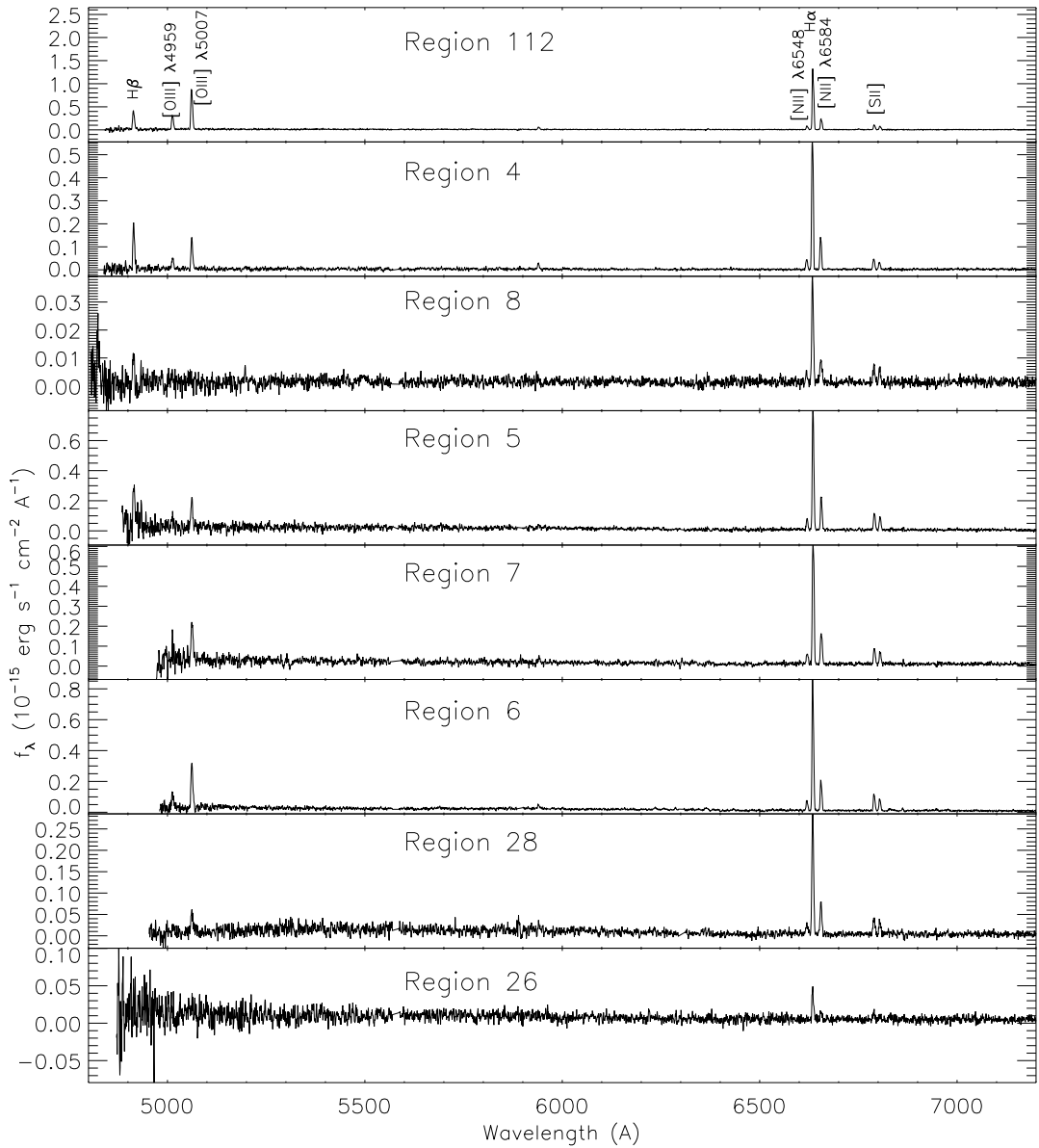}
\caption{Gemini/GMOS spectra of the star-forming regions located at the tip of the tidal tail of NGC 92. In this figure we show the spectra of regions that are shown in the bottom close-up of Fig. \ref{ngc92_rgb_regions_phot_paper_2}, starting from the northern to the southern regions. The spectra of sources TF7, TF6 and TF28 did not cover the H$\beta$ emission line. Some interpolation feature may be seen when gap and skyline artefacts were removed with the procedure presented in \S 2.1.2.}
\label{plot_spectra3}
\end{figure*}

\begin{table}
\centering
\begin{minipage}[t]{\columnwidth}
\caption{Physical properties of the star-forming regions located in the tidal tail of NGC 92.}
\begin{tabular}{ccccc}
\hline
System & (\textit{u\pr}-\textit{g\pr})\footnote{Magnitudes were measured in a fixed aperture of 0.5 arcsec radius and corrected by MW extinction and no internal extinction.} & (\textit{g\pr}-\textit{r\pr}) &  E(B-V)\footnote{Left values: Internal E(B-V) estimated from the photometry (see \S 3.2). Right values: Internal E(B-V) values derived from the spectroscopy (see \S \ref{line_fluxes}).} & Ages\footnote{Ages estimated from (\textit{u\pr}-\textit{g\pr}) and (\textit{g\pr}-\textit{r\pr}). In parenthesis we show the ages estimated from the H${\alpha}$ equivalent width and SB99 models.} \\
   &  mag  & mag &  mag  &  Myr  \\
\hline
  TF1 &  0.77$\pm$0.03  & 0.25$\pm$0.02 & 0.34/0.10 & 7$\pm^{7}_{7}  $ (5) \\ 
  TF2 &  0.79$\pm$0.04  & 0.53$\pm$0.02 & 0.62/0.42 & 6$\pm^{5}_{6}  $ (5) \\
  TF4 &  0.51$\pm$0.07  & 0.68$\pm$0.06 & 0.78/0.38 & 1$\pm^{5}_{1}  $ (1) \\
  TF5 &  0.69$\pm$0.04  & 0.63$\pm$0.02	& 0.82/0.76 & 3$\pm^{2}_{3}  $ (5) \\
  TF6 &  0.69$\pm$0.03  & 0.37$\pm$0.03	& 0.28/0.55 & 8$\pm^{8}_{8}  $ (5) \\
  TF7 &  0.58$\pm$0.03  & 0.34$\pm$0.02	& 0.62/0.69 & 4$\pm^{4}_{4}  $ (5)  \\
  TF8 &  1.08$\pm$0.06  & 0.39$\pm$0.04	& 0.46/0.09 & 7$\pm^{7}_{7}  $ (6)  \\ 
 TF10 &  0.94$\pm$0.06  & 0.36$\pm$0.04	& 0.42/0.66 & 7$\pm^{7}_{7}  $ (6)  \\
 TF11 &  1.06$\pm$0.14  & 0.30$\pm$0.09	& 0.40/0.42 & 7$\pm^{7}_{7}  $ (7).\\
 TF26 &  0.99$\pm$0.10  & 0.34$\pm$0.06	& 0.40/0.61 & 7$\pm^{7}_{7}  $ (7).\\
 TF28 &  0.89$\pm$0.07  & 0.29$\pm$0.04	& 0.36/0.65 & 7$\pm^{7}_{7}  $ (5)  \\
 TF29 &  0.79$\pm$0.07  & 0.58$\pm$0.03	& 0.80/0.34 & 4$\pm^{3}_{4}  $ (5)  \\
TF112 &  0.76$\pm$0.02  & 0.44$\pm$0.01	& 0.70/0.65 & 4$\pm^{4}_{4}  $ (2)  \\ 
TF115 &  0.76$\pm$0.05  & 0.34$\pm$0.02	& 0.42/...  & 6$\pm^{6}_{6}  $ NS   \\
TF116 &  0.92$\pm$0.06  & 0.61$\pm$0.02	& 0.76/...  & 5$\pm^{5}_{5}  $ NS   \\
TF117 &  0.86$\pm$0.06  & 0.34$\pm$0.05	& 0.40/...  & 7$\pm^{7}_{7}  $ NS   \\
\hline
\vspace{-0.8cm}
\label{tablephot}
\end{tabular}
\end{minipage}
\end{table}

\subsection{Optical Radial Velocities, fluxes, extinctions and equivalent widths}
\label{line_fluxes}

In order to identify the spectral lines and to determine the radial velocities of the observed targets, we cross-correlated the spectra with a template containing all typical \hii region lines. For this purpose we applied the IRAF package EMSAO \citep{Kurtz98}. We note that the arc calibration resulted in spectra with rms wavelength dispersions better than 0.5 {\AA}. The resulting emission lines velocities and their uncertainties are given in Table \ref{table3}. 

Spectroscopic observations with a R400$\_$G5325 grating are far from ideal for accurate radial velocity studies, but combining the results of all identified emission lines in each spectrum it is possible to have a mean accuracy of \mbox{40 km s$^{-1}$.} 

As a result of the cross-correlation procedure, we also obtained a list of emission lines identified with emission levels higher than at least one sigma from the local rms spectra. The same procedure also produce a list of equivalent width for each spectral line. Using this list of emission lines, we integrate the flux under each spectral line considering the following models: (a) Gaussian; (b) Lorentzian; (c) numerical integration and (d) double Gaussian fitting. We calculated the mean between the two best procedures to measure fluxes, which were the Gaussian and the numerical integration, coherent with the method adopted by \cite{Zaritsky94} to analyze their data sample of HII regions. Our method has the advantage to allow us to register and identify profiles different than usual. No remarkable distortions were found in the observed profiles (details of this procedure can be found in \cite{Scarano11}). The equivalent width of each line was derived in the previous step using the package EMSAO.

Fluxes in spectral lines are affected differentially by galactic and intrinsic contributions according to the wavelength of the line. The galactic contribution is well established by \cite{Amores05} and we applied their model to correct all spectra for the galactic flux losses. Assuming that the remaining differences in the ratio of the Balmer lines are a consequence of the intrinsic extinction, we corrected each spectrum again for the reddening by determining the Balmer line ratios \ha and \hb, when both were available. The value for the intrinsic \ha/\hb ratio was taken from \cite{Osterbrock89} for $T_e$=10000 K and $N_e$=100, using the Galactic extinction law and the IDL code FM\_UNRED based on \cite{Fitzpatrick99}. In reason of the differences in the spectral coverage along the observed GMOS field, a few spectra did not present \hb. For these cases we adopted the extinction to be the median value of the extinction in the three nearest regions for which it was possible to calculate the \ha/\hb ratio (see Table \ref{tablephot}). The corrected emission-line fluxes measured for all the \hii regions belonging to NGC 92 are presented in Table \ref{tablefluxes}. No significant underlying Balmer absorption was identified in the spectra, so no corrections of this order were applied.

In order to determine the spectroscopic age of the star-forming regions, we used the equivalent width of the H$\alpha$ emission line of each spectrum. We then compared this value with the equivalent width derived from the SB99 model, for a single stellar population and a Salpeter IMF.

\begin{figure}
\includegraphics[width=1.05\columnwidth]{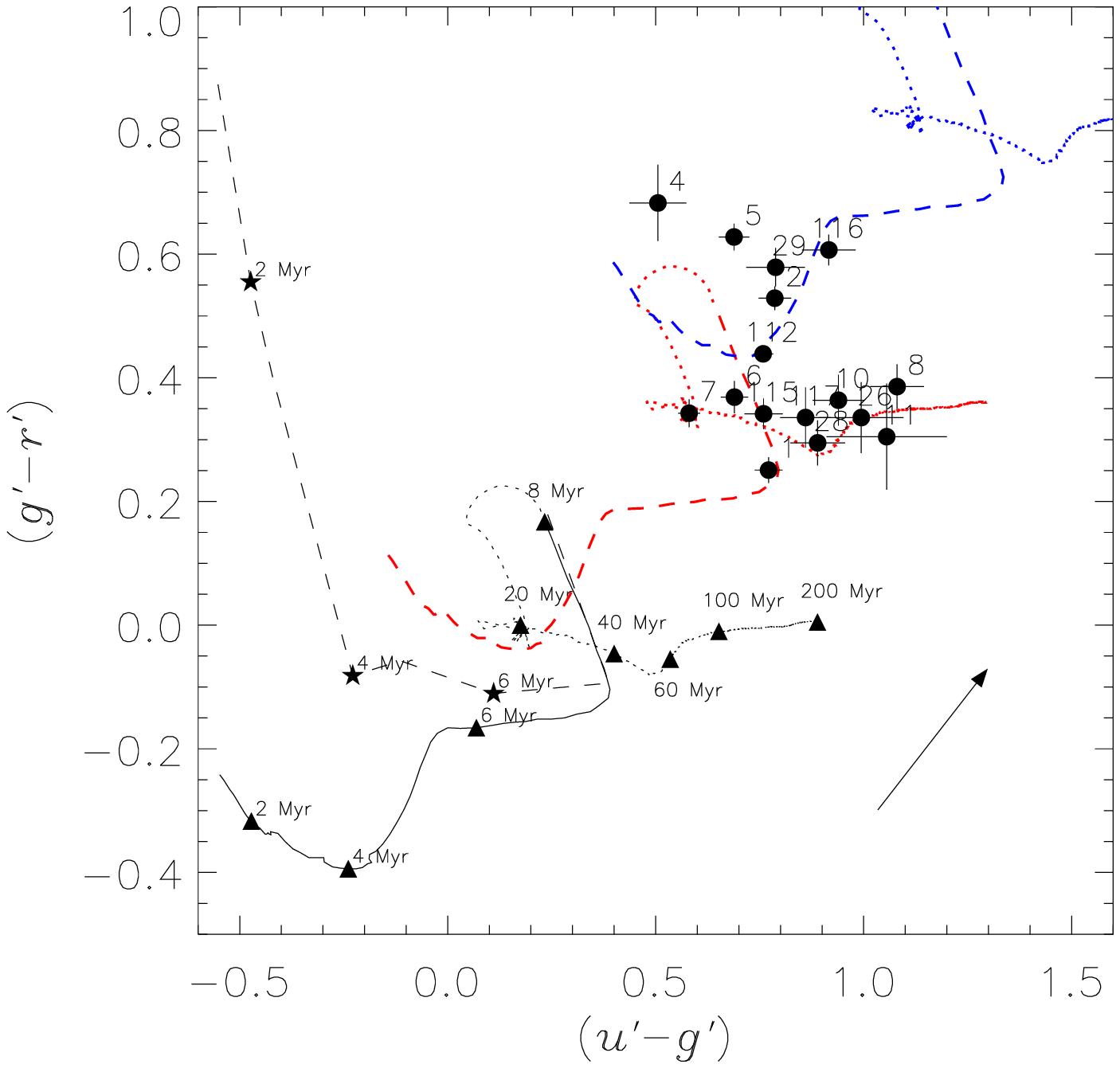}
\caption{Gemini (\textit{u\pr}-\textit{g\pr}) versus (\textit{g\pr}-\textit{r\pr}) diagram for the star-forming regions located in the tidal tail of NGC 92. On the same diagram we have overplotted the Starburst99 models tuned for the Gemini bands. No internal extinction is represented by a black line, where the black dashed line represents the contribution of the H$\alpha$ emission line in the broadband colours (taken from \citealt{Smith08}) and the dotted black line displays the SSP models for ages larger than 8 Myrs (not used in the age determination given the emission line nature of our sources). Solid black triangles mark the different ages displayed by the models. The red and blue dashed lines represent a starburst extinction law (Calzetti et al. 1994) with an E(B-V)=0.3 (A$_{V}$=1.2 mag) and E(B-V)=0.7 (A$_{V}$=2.8 mag), respectively (where we have used \mbox{$A_{V}$= E(B-V) $\times$ 4.0).}}
\label{gr_ug}
\end{figure}

\begin{table*}
\centering
\begin{minipage}[t]{\textwidth}
\scriptsize
\caption{Line fluxes for the star-forming regions located in the tidal tail and main body of NGC 92}
\begin{tabular}{ccccccccc}
\hline
ID  & [OIII]$\lambda$4363 & H$\beta$\footnote{The spectral coverage did not cover the H$\beta$ emission line for regions 2, 6, 7, 13, 16, 17, 18, 28, 29, 36, 44, 47 and 106.}  & [OIII]$\lambda$4959 & [OIII]$\lambda$5007 & [NII]$\lambda$6548 & H$\alpha$ & [NII]$\lambda$6584  & [ArIII]$\lambda$7136 \\
\multicolumn{9}{c}{10$^{-16}$ erg cm$^{-2}$ s$^{-1}$ $\rm\AA^{-1}$} \\
\hline
TF1   &  --  &    2.02$\pm$0.47  &     0.48$\pm$0.16  &    2.44$\pm$0.51  &     0.34$\pm$0.11  &     6.25$\pm$0.10  &     1.59$\pm$0.02   &   --   \\  
TF2   &  --  &    --  &     2.51$\pm$0.72  &   11.40$\pm$0.14  &     1.88$\pm$0.64  &    26.00$\pm$0.50  &     6.10$\pm$0.14   &   1.09$\pm$0.31   \\  
TF4   &    0.47$\pm$0.39  &   12.00$\pm$0.93  &	 2.87$\pm$0.78  &    8.19$\pm$0.45  &	 2.20$\pm$0.96  &    29.00$\pm$0.52  &     7.22$\pm$0.14   &   0.46$\pm$0.12   \\  
TF5   &   --  &   26.40$\pm$8.23  &	 4.91$\pm$3.03  &   17.60$\pm$4.49  &	 6.19$\pm$2.37  &    79.60$\pm$1.19  &    21.70$\pm$0.21   &   0.97$\pm$0.38   \\  
TF6   &  --  &    --  &	 9.82$\pm$2.76  &   35.30$\pm$1.09  &	 5.80$\pm$1.30  &    82.00$\pm$0.97  &    19.60$\pm$0.24   &   1.58$\pm$0.40   \\  
TF7   &   --  &    --  &	 7.13$\pm$4.01  &   20.50$\pm$4.76  &	 5.19$\pm$2.18  &    68.50$\pm$1.43  &    18.20$\pm$0.29   &   1.39$\pm$0.49   \\  
TF8   &   --  &   10.60$\pm$0.73  &	 2.12$\pm$0.54  &    6.49$\pm$0.49  &	 --  &     --  &     --   &   --   \\  
TF10  &  --  &    7.73$\pm$3.08  &	 0.20$\pm$0.14  &   12.20$\pm$4.16  &	 1.45$\pm$0.87  &    23.40$\pm$0.37  &     8.15$\pm$0.30   &   0.33$\pm$0.24   \\  
TF11  &    --  &    0.47$\pm$0.34  &	 --  &    --  &	 --  &     0.34$\pm$0.11  &     --   &   --   \\  
TF12  &   --  &    7.88$\pm$1.78  &	 2.14$\pm$1.13  &    6.50$\pm$1.49  &	 1.96$\pm$0.72  &    25.00$\pm$0.61  &     7.29$\pm$0.11   &   --   \\  
TF15  &    --  &   18.10$\pm$4.84  &	 4.08$\pm$1.29  &   14.90$\pm$0.63  &	 5.06$\pm$1.87  &    56.10$\pm$1.43  &    16.60$\pm$0.29   &   1.03$\pm$0.32   \\  
TF16  &   --  &    --  &	11.10$\pm$2.63  &   35.00$\pm$0.61  &	 3.96$\pm$0.69  &    70.40$\pm$1.28  &    15.30$\pm$0.50   &   1.55$\pm$0.06   \\  
TF17  &    --  &    --  &	10.20$\pm$0.48  &   31.60$\pm$0.19  &	 0.86$\pm$0.14  &    24.00$\pm$0.09  &     3.32$\pm$0.07   &   0.84$\pm$0.22   \\  
TF18  &    --  &    --  &	 --  &    1.05$\pm$0.84  &	 0.66$\pm$0.33  &     8.26$\pm$0.11  &     2.99$\pm$0.12   &   --   \\  
TF26  &    --  &    0.56$\pm$0.18  &	 0.67$\pm$0.36  &    0.25$\pm$1.08  &	 1.90$\pm$0.45  &     3.80$\pm$0.91  &     1.18$\pm$0.48   &   0.06$\pm$0.02   \\  
TF28  &    --  &    --  &	 0.78$\pm$0.36  &    2.56$\pm$0.74  &	 0.50$\pm$0.16  &    15.00$\pm$0.31  &     3.88$\pm$0.88   &  -0.01$\pm$0.08   \\  
TF29  &    --  &    --  &	 0.19$\pm$0.07  &    0.86$\pm$0.27  &	 0.31$\pm$0.18  &    13.50$\pm$0.20  &     3.38$\pm$0.77   &   --   \\  
TF36  &    --  &    --  &	34.90$\pm$0.82  &  107.00$\pm$0.31  &  229.00$\pm$57.70  &  2570.00$\pm$86.00  &  1430.00$\pm$5.25   &  18.70$\pm$0.40   \\  
TF47  &    --  &    --  &	 --  &    --  &	 --  &     0.28$\pm$0.10  &     0.09$\pm$0.06   &   --   \\  
TF112 &   --  &   43.20$\pm$0.61  &	34.30$\pm$1.03  &   95.00$\pm$0.40  &	 7.39$\pm$2.26  &   128.00$\pm$1.69  &    24.60$\pm$0.56   &   4.02$\pm$0.17   \\  
\hline
\vspace{-0.8cm}
\label{tablefluxes}
\end{tabular}
\end{minipage}
\end{table*}

\subsection{Radial distance estimaties}
\label{distances}

In order to determine the metallicity gradient for NGC 92, it is necessary to estimate the distance from each star-forming region to the center of this galaxy. Traditional procedures to calculate distances rely on the determination of the projection parameters of a galaxy based on isophotes, velocity fields or the geometry of spiral arms. However, all these components may be distorted due to tidal interactions. For NGC 92 the velocity field reveals a disk-like rotation, with a wide radial velocity range and a relative symmetric rotation curve, at least inside the first 40 arcsec \citep{TorresFlores09}, near the R$_{25}$ level (R$_{25}$$\sim$10 kpc) in the B-filter \citep{Lauberts89}. This feature suggests that, in spite of the interactions, the central parts of this galaxy is still dominated by the angular momentum and therefore the hypothesis based in the disc symmetry is still good to determine the projection parameters of NGC 92 (we note that \citealt{Bastian09} found a similar disk-like rotation for several star clusters located in the Antennae merging system). Based on the velocity field of NGC 92, \cite{TorresFlores10a} have determined the kinematic inclination and position angle as 48\dg and 146\dg, respectively. Morphology studies using the isophotes agree with the position angle (149\dg), but the inclination are appreciably larger (62\dg - 70\dg - see references in Table 3 in \citeauthor{TorresFlores10a}). Since no clear spiral structure was detected it was not possible to obtain the projection parameters with these method. However, using the morphological procedures described by \cite{Sca2008a}, in which we masked the contribution of stellar fields, the interarm regions and the tidal tail, we obtained an inclination of 51\dg and a position angle of 155\dg. These results seem to be more coherent with the kinematic results, and are also representative of the morphology observed in the central region of NGC 92. For these reasons we adopted these values to calculate the distances following the expressions described by \cite{Sca2008a}. To convert angular to linear distances, we used the distance to NGC 92 (43.2 Mpc, \citealt{Karachentsev96}).

\subsection{Oxygen Abundances}
\label{oxygen_abundances}

Given that our spectra do not cover the region of the temperature sensitive [O~{\sc iii}]$\lambda$4363 {\AA} line, we estimate the oxygen abundances using semi-empirical methods. The calibrators applied were the [N~{\sc ii}]/$H\alpha$ method (hereafter N2, \citealt{PP04}, where we adopt a cubic fit on the data) and the [O~{\sc iii}]/[N~{\sc ii}] method (hereafter O3N2, \citealt{PP04} and \citealt{Stasinska06}) and the [Ar~{\sc iii}]/[O~{\sc iii}] method of \cite{Stasinska06}. The observational setup was optimized to observe the [N~{\sc ii}] and $H\alpha$ lines, and for this reason most of the procedures could not be applied for all spectra. Table \ref{tableabundances} summarizes the results for the oxygen abundances.

The uncertainties in the metallicities were calculated by propagating the flux uncertainties, which was assumed to be the standard deviation of the flux derived in the multi-profile fitting procedure (see \S \ref{line_fluxes}). In general the results are in agreement with uncertainties obtained in the literature (see \citealt*{Pilyugin2004} and references therein) however the results obtained with N2 method are clearly underestimated, because the dispersion in the line fluxes for different profiles do not result in significant fluxes differences. By taking into account a term obtained by propagating the uncertainty in extinction, an extra component of 0.08 dex may be added to the resulting uncertainties of the N2 metallicities, with results comparable to those by \cite{Bresolin12}.

One should note that the flux uncertainties were estimated by accounting the uncertainties in the emission lines fluxes, the photon statistics, and the extinction. Uncertainties related with the calibration methods are much larger than the gradients themselves (of the order of 0.2 dex). However, as pointed out in \cite{Scarano11} and \cite{Bresolin12} different calibration methods may deliver different metallicities levels, but the gradients are less affected by these uncertainties, so differential analysis methods are robust and the choice of different calibrations do not affect the slopes significantly.

\begin{table*}
\centering
\begin{minipage}[t]{\textwidth}
\scriptsize
\caption{Oxygen abundances for the star-forming regions located in the tidal tail and main body of NGC 92}
\begin{tabular}{cccccccc}
\hline
ID & 
12+log(O/H)\footnote{Oxygen abundances estimated by using the N2 calibrator proposed by \citep{Pettini04} and using a cubic fit.} & 
12+log(O/H)\footnote{Oxygen abundances estimated by using the O3N2 calibrator proposed by \citep{Pettini04}.} & 
12+log(O/H)\footnote{Oxygen abundances estimated by using the O3N2 calibrator proposed by \citep{Stasinska06}.} & 
12+log(O/H)\footnote{Oxygen abundances estimated by using the [ArIII]/[OIII] calibrator proposed by \citep{Stasinska06}.} & 
E(B-V)\footnote{Internal extinction for each source. These values were estimated following \citep{Fitzpatrick90} } & 
EW\footnote{Equivalent width of the H$\alpha$ line.}&
Distance\footnote{Distance from the center of the galaxy NGC 92} \\
   &   &  &    &  & mag & {\AA} & kpc \\

\hline
TF1   &   8.54$\pm$0.10 &  8.49$\pm$0.19  &  8.36$\pm$0.19 &  --                          & 0.10$\pm$0.06  & 235     &14.19\\
TF2   &   8.51$\pm$0.10 &  --                           &  --                          & 8.59$\pm$0.26  & 0.42$\pm$0.01  & 333     & 10.1 \\
TF4   &   8.53$\pm$0.10 &  8.55$\pm$0.19  &  8.41$\pm$0.19  & 8.43$\pm$0.26  & 0.38$\pm$0.03  & 2271   & 18.7 \\
TF5   &   8.57$\pm$0.10 &  8.57$\pm$0.19  &  8.43$\pm$0.19  & 8.44$\pm$0.26  & 0.76$\pm$0.13  & 484     & 22.72\\
TF6   &   8.52$\pm$0.10 &  --                           &  --                          & 8.37$\pm$0.26  & 0.55$\pm$0.03  &  346    & 25.06\\
TF7   &   8.56$\pm$0.10 &  --                           &  --                          & 8.49$\pm$0.26  & 0.69$\pm$0.13  & 317     & 24.15\\
TF8   & --                         &  --                           &  --                          & --                           & 0.09$\pm$0.00  & 143     & 21.74\\
TF10  & 8.67$\pm$0.10&  8.53$\pm$0.19   &  8.39$\pm$0.19 & 8.27$\pm$0.26  & 0.66$\pm$0.16  & 168     & 13.07\\ 
TF11  & --                         &  --                           &  --                          & --                           & 0.42$\pm$0.02   &   35     & 11.5 \\
TF12  & 8.60$\pm$0.10 &  8.55$\pm$0.19  &  8.41$\pm$0.19 & --                           & 0.55$\pm$0.03   &  210    & 6.17 \\
TF15  & 8.60$\pm$0.10 &  8.55$\pm$0.19  &  8.41$\pm$0.19  & 8.51$\pm$0.26  & 0.66$\pm$0.09   & 228    & 4.33 \\
TF16  & 8.48$\pm$0.10 &  --                           &  --                          & 8.36$\pm$0.26  & 0.49$\pm$0.03    & 261    & 8.05 \\ 
TF17  & 8.35$\pm$0.10 &  --                           &  --                          & 8.13$\pm$0.26  & 0.38$\pm$0.03   & 2538 & 9.97 \\
TF18  & 8.69$\pm$0.10 &  --                           &  --                          & --                           & 0.56$\pm$0.03  & 88       & 6.46 \\
TF26  & --                         &  8.50$\pm$0.17   &  8.37$\pm$0.19 & 8.55$\pm$0.26  & 0.61$\pm$0.03   & 45      & 27.23\\ 
TF28  & 8.55$\pm$0.10&  --                           &  --                          & --                           & 0.65$\pm$0.03   & 345    & 25.7 \\
TF29  & 8.54$\pm$0.10&  --                           &  --                          & --                           & 0.34$\pm$0.03   & 262    & 8.94 \\ 
TF36  & --                        &  --                           &  --                           & 8.68$\pm$0.26  & 0.39$\pm$0.06   & 184    & 0.25 \\
TF47  & 8.63$\pm$0.10&  --                           &  --                           & --                           & 0.39$\pm$0.06    & --       & 11.29\\
TF112 & 8.45$\pm$0.10 &  8.35$\pm$0.19  &  8.25$\pm$0.19  & 8.33$\pm$0.26   & 0.65$\pm$0.01    & 1294 & 17.63\\
\hline
\vspace{-0.8cm}
\label{tableabundances}
\end{tabular}
\end{minipage}
\end{table*}

\section{Results}

In this section, we list the main photometric and spectroscopic results obtained for the star-forming regions located in the main body and the tidal tail of NGC 92.

\subsection{A general view of NGC 92}

In Fig. \ref{ngc92_rgb_regions_phot_paper_2} we show a Gemini/GMOS optical image of NGC 92. Regions of interest in this study (and for which we have spectra) are labeled by numbered red circles. Inspecting this image, it results clear the presence of an extended tidal tail \citep[$\sim$30 kpc][]{Temporin05}, which displays several compact knots that were reported as H$\alpha$ emitting sources. These regions are shown in a close-up at the left panels of Fig. \ref{ngc92_rgb_regions_phot_paper_2}. At the bottom left panel of Fig. \ref{ngc92_rgb_regions_phot_paper_2}, we can see the tidal dwarf galaxy (TDG) candidate previously found by \citep{TorresFlores09}, based on its age and location (which is placed at the tip of the tidal tail). In the Gemini/GMOS images this object shows diffuse emission and several blue star-forming knots. In Figs. \ref{plot_spectra1}, \ref{plot_spectra2} and \ref{plot_spectra3} we plot the spectrum of the star-forming complexes detected in the main body and tidal tail of NGC 92. In all cases, the most intense emission lines correspond to H$\alpha$, H$\beta$, [OIII] $\lambda$5007 $\rm\AA$ and [NII]$\lambda$6584 $\rm\AA$. We note that regions located in the tidal tail display typical spectrum of \hii regions.

\subsection{Colors, ages and internal extinctions}

In Fig. \ref{gr_ug} we plot the (\textit{u\pr}-\textit{g\pr}) and (\textit{g\pr}-\textit{r\pr}) colors for the regions located in the tidal tail of NGC 92, where we have included the SB99 models with no extinction (black line) and the models derived for the Calzetti et al. (1994) extinction law and color excess of \mbox{E(B-V)=0.3 mag} and \mbox{E(B-V)=0.7 mag}, in red and blue, respectively. In this Figure, the black dashed line represent the nebular contribution of the H$\alpha$ emission line in the broadband colours, which was taken from \cite{Smith08}. This contribution is  important during the first million years of a star-forming region, specially in their emission in the \textit{g\pr} and \textit{r\pr}-band filters (see \citealt{Smith08}). Given that most of the sources are located in a region of the plot where the H$\alpha$ contribution is not significantly important (with the exception of TF4), the photometric ages were estimated directly from the SB99 models. As a comparison, in this Figure we have also included the models for ages larger than 8 Myrs, which was the limit in the age determination, given the spectra of the analysed sources. In Table \ref{tablephot} we list the object ID, (\textit{u\pr}-\textit{g\pr}) and (\textit{g\pr}-\textit{r\pr}) colors, internal extinction and ages (estimated from the photometry) that we derived for these regions. For comparison, in this table we have included the internal extinctions, \mbox{E(B-V)}, and the ages that we derived from the spectroscopic observations of each source, as discussed in \S \ref{line_fluxes} (spectroscopic internal extinctions with their uncertainties are listed in Table \ref{tableabundances}). We found that the star-forming regions located in this tail display ages ranging from $\sim$1 to $\sim$8 Myrs, as expected from the used models (see \S 3.2). We found that ages derived from the H$\alpha$ equivalent widths are in agreement with the values derived just from the photometry. These young ages suggest that these clumps were formed \textit{in situ}.

In the case of the photometrically derived internal extinctions, we found that the regions located in the tail of NGC 92 display values ranging from E(B-V)$\sim$0.28 mag to E(B-V)$\sim$0.80 mag, with a mean of 0.53 mag. In the case of the spectroscopic values, they span from E(B-V)$\sim$0.09 mag to E(B-V)$\sim$0.76 mag, with a mean of 0.49 mag. On average, these values are higher than the extinctions reported by regions located in other tails, for instance, NGC 4017, NGC 2856 and NGC 2782 (\citealt{Hancock09}, \citealt{Smith08} and \citealt{TorresFlores12}, respectively), which report average values of 0.14, 0.13 and 0.24 magnitudes. However, the extinctions found in the tail of NGC 92 are in agreement with the extinction found in the TDG located at the tip of the tidal tail of Arp 245 (E(B-V)=0.6, \citealt{Duc00}).

\begin{figure}
\includegraphics[width=\columnwidth]{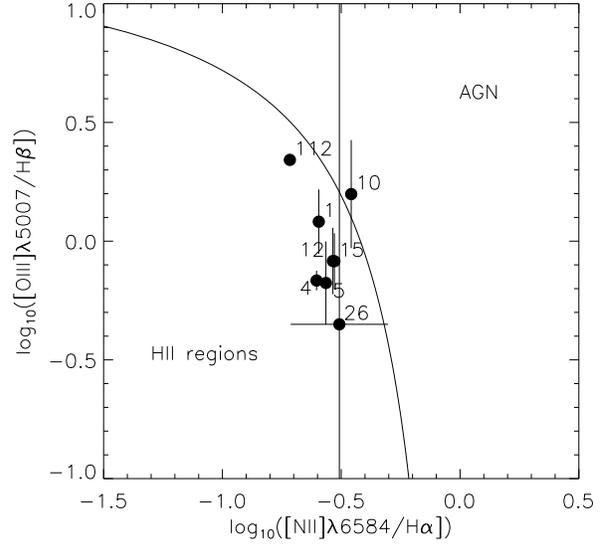}
\caption{BPT diagram for the star-forming regions located in NGC 92. The continuous line corresponds to the limit between \hii regions and AGNs, as described in Kauffmann et al. (2003).}
\label{BPT}
\end{figure}

We note that sources TF8 and TF11 are slightly off the region defined by the models used in the age/extinction determination (models ranging from 1 to 8 Myr). In fact, Figure \ref{gr_ug} suggests that these sources can be explained with models having ages larger than 90 Myrs. However, TF8 displays strong emission lines in its spectrum, suggesting the presence of a young stellar population. The case of TF 11 seems to be slightly different. The spectrum of this source presents a weaker H$\alpha$ emission line. Therefore, under the assumption of an instantaneous burst, source TF11 can not have an age larger than $\sim$10 Myr. Given that, TF8 and TF11 have experienced a recent burst of star formation and their position in the color-color plot should be explained for other mechanism. For example, a different extinction law could explain the current position of these sources in this diagram, given that we assumed a starburst extinction law.

\subsection{Line fluxes and oxygen abundances}

In Table \ref{tablefluxes} we list the main line fluxes that we obtain from the spectrum of the star-forming regions located in the tidal tail and the main body of NGC 92. In order to confirm the ionisation mechanism of the sources, we have used the fluxes in the [OIII]$\lambda$5007 {\AA}, H$\beta$, H$\alpha$ and [NII]$\lambda$6584 {\AA} emission lines to analyse the BPT diagram (\citealt{Baldwin81}), as shown in Fig. \ref{BPT}. In this figure, the solid line corresponds to the limit between \hii regions and active galactic nucleus (AGNs), as defined by \cite{Kauffmann03}. Taken into account the uncertainties in the flux measurements, we found that all objects lie in the \hii region locus, discarding shocks as a mechanism in the ionization of the gas.

Fluxes were used to estimate the oxygen abundance of each star-forming region (see \S \ref{oxygen_abundances}), which are listed in Table \ref{tableabundances}. Depending on the empirical calibrator (and the available emission lines), we list up to four values for the oxygen abundances. In the following, when we refer to the oxygen abundance of a specific star-forming region, we refer to the value derived by using the N2 method and the cubic fit given in \cite{Pettini04}, given that this method was useful for most of the regions. In the case of the central region of NGC 92, we will use the [ArIII]/[OIII] calibrator given the limitation in the use of the N2 method \citep{Stasinska06}. However, given the weakness of the [ArIII] line, we will use this value as a crude estimation of the central metallicity of NGC 92. In Table \ref{tableabundances} we also list the internal extinction for each system, as discussed in \S \ref{line_fluxes}. The distance from the center of NGC 92 to each star-forming complex is also recorded. Inspecting the values listed in Table \ref{tableabundances} we found that most of the regions in NGC 92 have slightly sub-solar oxygen abundances (taking \mbox{12+log(O/H)$_{\odot}$=8.72} as the solar value, \citealt{AllendePrieto01}), with values ranging from 12+log(O/H)=8.35 to 12+log(O/H)=8.69. The center of NGC 92 has an oxygen abundance of $12+\rm{log(O/H)}=8.68\pm0.25$. Taking into account the uncertainties, this value is similar than the abundances displayed by the central regions of the galaxy pairs studied by \cite{Kewley10}, which display values ranging from $12+\rm{log(O/H)}=8.75$ to $12+\rm{log(O/H)}=9.08$ (these pairs present a separation between 15 and 25 kpc and M$_{B}<-20$). In the case of the regions located in the tail of NGC 92, they display oxygen abundances that span from 12+\rm{log(O/H)}=8.45$\pm$0.10 to 12+\rm{log(O/H)}=8.67$\pm$0.10 (using the N2 calibrator). This range in abundances appears to be small, given the large extent of this tidal tail ($\sim$30 kpc). The oxygen abundances that we derived for the star-forming regions located in it are in agreement with those computed for other complexes formed in these kinds of structures (e. g. \citealt{Trancho12}). However, the abundances in the tail of NGC 92 are lower than the values found by \cite{Chien07} for the northern tidal tail of the interacting pair NGC 4676 ($12+\rm{log(O/H)}=8.92-8.97$), which is in an early stage of merging, where the tails have a dynamical age of $\sim$170 Myrs (\citealt{Chien07}), roughly consistent with the age of the tail of NGC 92. This fact shows the diversity in the chemical properties of newly formed objects in tidal tails/debris.

\begin{figure*}
\includegraphics[width=0.9\textwidth]{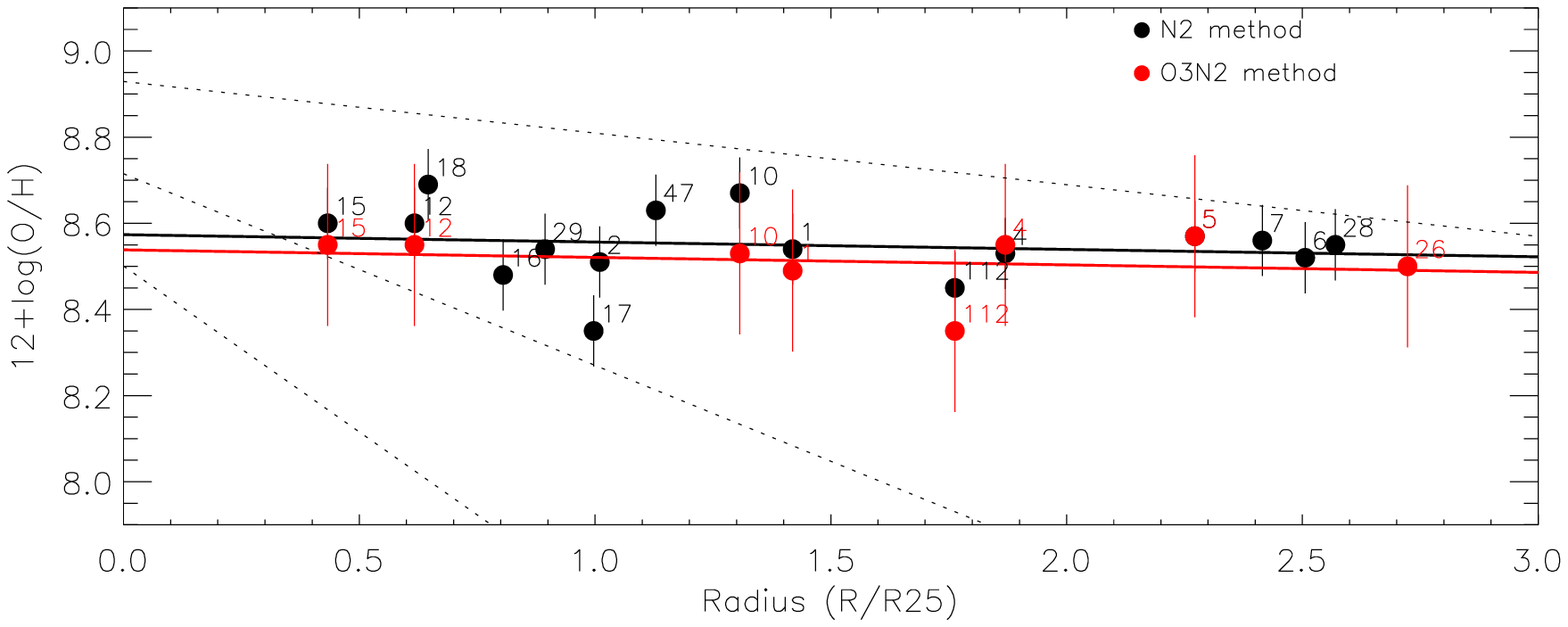}\\
\includegraphics[width=0.9\textwidth]{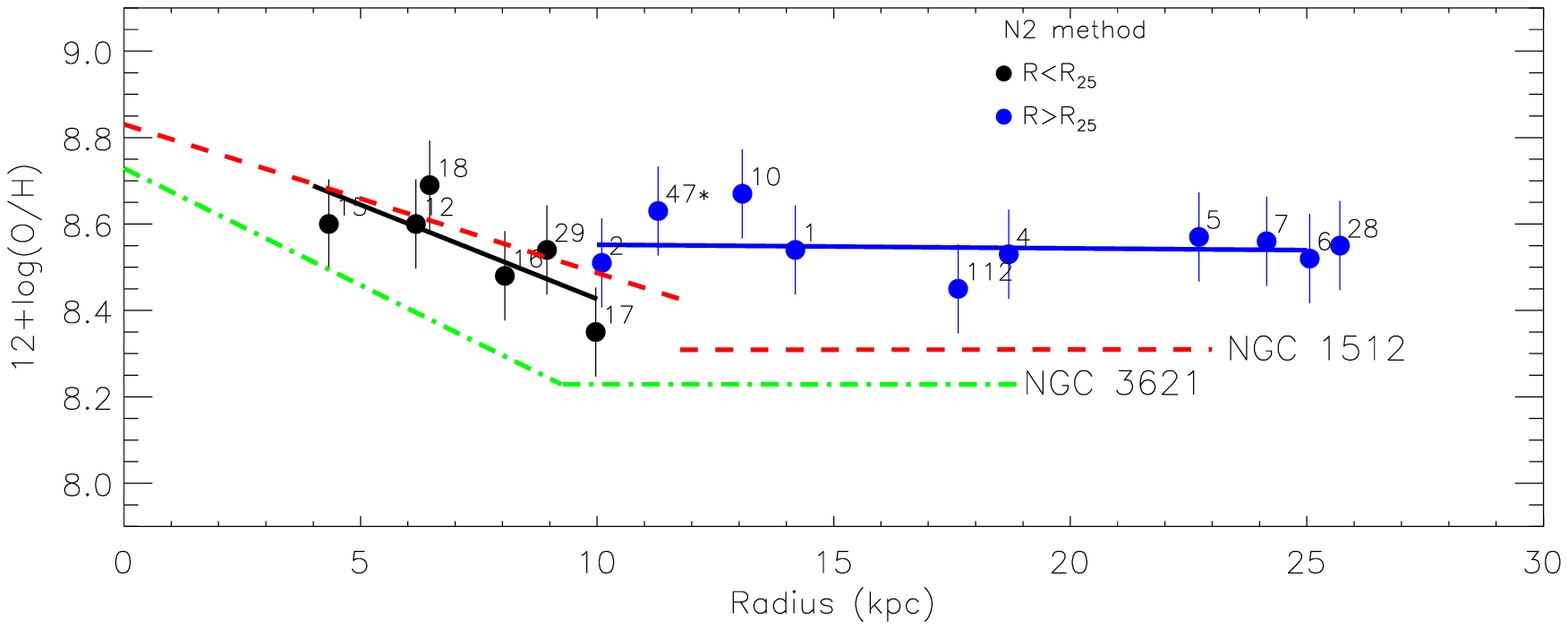}
\caption{Top panel: Radial Oxygen distribution for the star forming regions located in the whole galaxy NGC 92. Black and red filled circles indicate the abundances estimated by using the N2 and O3N2 methods, respectively. Black and red solid lines represent a linear fit on each data set. In this panel, the radius has been normalised to the optical radius of NGC 92. The dotted lines represent an average of the metal distribution of a sample of spiral galaxies studied by Pilyugin et al. (2004) (see \S 4.4). Bottom panel: Radial Oxygen distribution for NGC 92. Black filled circles represent regions located in the inner part of NGC 92 R$<$R$_{25}$ and blue filled circles correspond to regions located at radii larger than R$_{25}$. Black and blue solid lines represent linear fits on each data set. In the case of the outer region, the source TF47 was not included in the fit, given that this source is not located in the tidal tail of NGC 92. In the same panel we have plotted the results obtained by Bresolin et al. (2012) for the galaxies NGC 1512 and NGC 3621 (red dashed and green dotted dashed-lines, respectively).}
\label{MetalGradient}
\end{figure*}

\begin{figure*}
\includegraphics[width=0.9\textwidth]{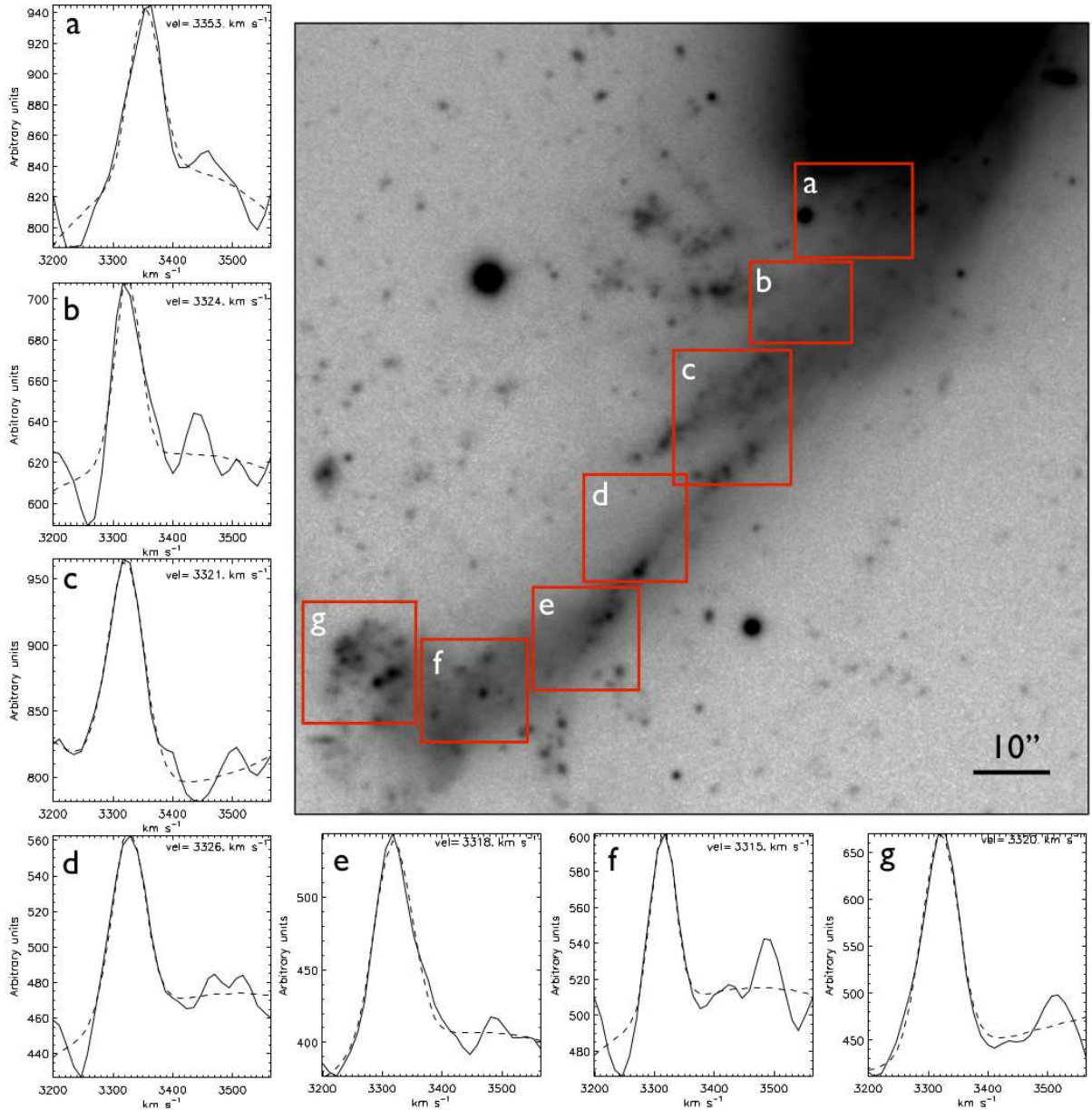}
\caption{H$\alpha$ profiles across the tidal tail of NGC 92 (taken from the Fabry-Perot data cube published in \citealt{TorresFlores09}, which has a pixel scale of 0.405 arcsec pixel$^{-1}$). In each panel is indicated the velocity of the profile, derived from a single Gaussian fitting.}
\label{fabryperot}
\end{figure*}

\begin{table}
\centering
\begin{minipage}[t]{\columnwidth}
 \caption{Zero points and slopes of the linear fits}
 \label{slopes}
\begin{tabular}{lcc}
\hline
Parameter & Zero point ($\alpha$) &  Slope ($\beta$)\\
 & 12+log(O/H) &  dex kpc$^{-1}$\\
\hline
\vspace{0.1cm}
NGC 92 (whole galaxy): & &\\
$[NII]/H\alpha$   &  8.57$\pm$0.05  & -0.0017$\pm$0.0021 \\
\vspace{0.1cm}
$[OIII]/[NII]$         &  8.54$\pm$0.02  & -0.0017$\pm$0.0017  \\
\hline
\vspace{0.1cm}
NGC 92 (R$<$R$_{25}$): & &\\
\vspace{0.1cm}
$[NII]/H\alpha$   &  8.86$\pm$0.12  & -0.0436$\pm$0.0158   \\
\hline
\vspace{0.1cm}
NGC 92 (R$>$R$_{25}$,$\sim$tail): & &\\
\vspace{0.1cm}
$[NII]/H\alpha$   &  8.56$\pm$0.07   & -0.0008$\pm$0.0032   \\
\hline
\vspace{-0.8cm}
\end{tabular} 
\end{minipage}
\end{table}

\subsection{The oxygen abundance gradient in NGC 92}
\label{oxygen_gradient}

We have obtained spectra for a large number of emitting-line regions in the tidal tail of NGC 92. This fact allowed us to study in detail the oxygen abundance gradient along this structure and compare it with the abundances displayed by the central regions of NGC 92. In the top panel of Fig. \ref{MetalGradient} we plot the oxygen abundance for each star-forming region in the main body and the tail of NGC 92 versus the distance to the center of this system, where the distance was corrected by the inclination and position angle of the galaxy (see \S \ref{distances}). In this panel, the distances has been normalised by the optical radius of the galaxy ($\sim$10 kpc). Black and red filled circles represent the oxygen abundances estimated by using the N2 and O3N2 empirical calibrators, as shown in Table \ref{tableabundances} and the continuous solid lines represent a linear fit to each data set. In Table \ref{slopes} we list the zero point and the slopes of both fits. The top panel of Fig. \ref{MetalGradient} shows a small difference in the zero point of the different fits ($\alpha$=8.57$\pm$0.05 for the N2 method versus $\beta$=8.54$\pm$0.02 for the O3N2 method). This fact is derived from the different calibrators we used in the oxygen abundance estimation. Given that this difference is systematic, our analysis of the metallicity gradients are reliable when the same calibrator is used \citep[see][]{Kewley10}. In this Figure, the dotted black lines represent an average estimation for the metal distribution of a sample of spiral galaxies analysed by \cite{Pilyugin2004}, where the upper and lower dotted lines display 1$\sigma$ in zero point and slope. We note that most of the metal distributions studied by \cite{Pilyugin2004} included regions located inside the optical radius of the galaxies, and therefore, the dotted lines shown by us after R$_{25}$ are extrapolations of our average fits.

Inspecting the top panel of Fig. \ref{MetalGradient}, we can note that most of the regions in NGC 92 presents a similar oxygen abundance (for a given calibrator), which produces an almost flat metallicity gradient. In the case of the N2 method, we found a slope of $\beta$=-0.0017$\pm$0.0021, while in the case of the O3N2 method, we found a value of $\alpha$=-0.0017$\pm$0.0017 dex kpc$^{-1}$. We note that in both cases we are including regions located in the main body and in the tidal tail of NGC 92.

Given that the optical tidal tail of NGC 92 starts at $\sim$10 kpc (where 10 kpc corresponds to the optical radius of NGC 92, R$_{25}$=10 kpc, \citealt{TorresFlores09}), we can derive a linear fit for regions located at R$<$R$_{25}$ and R$>$R$_{25}$, in a similar way to what was done by \cite{Bresolin12} for the extended disks of NGC 1512 and NGC 3621. In the case of regions located at R$>$R$_{25}$, we are basically obtaining the oxygen abundance gradient of the tidal tail. One only exception is TF47 which is not in the tidal tail, but instead it is to the north of the galaxy. For this reason TF47 is not included in the linear fit for R$>$R$_{25}$. In the bottom panel of Fig. \ref{MetalGradient} we show both fits, where the black line indicates the linear fit for regions located at R$<$R$_{25}$ and the blue line corresponds to the fit over regions located at R$>$R$_{25}$ (the zero points and slopes for both fits are listed in Table \ref{slopes}). Despite the low number of regions located at R$<$R$_{25}$, the slope of the abundance gradient of the inner region of NGC 92 ($\beta$=-0.0436$\pm$0.0158) is different from the slope shown by the outer region of this galaxy ($\beta$=-0.0008$\pm$0.0032). In fact, the slope of the outer region of NGC 92 (R$>$R$_{25}$) is similar to the slope of the whole galaxy (when the N2 method is used and the uncertainties are considered). For comparison, in the bottom panel of Fig. \ref{MetalGradient} we show the oxygen radial distribution of the galaxies NGC 1512 (red dashed line) and NGC 3621 (green dotted-dashed line), for which the oxygen abundances were estimated from the N2 calibrator \citep[see][]{Bresolin12}. In the case of the inner region of NGC 92, we found a good agreement with the gradients derived by \cite{Bresolin12} for NGC 1512 and NGC 3621 (and also with the slope of the average metal distribution taken from the sample of spiral galaxies studied by  \cite{Pilyugin2004}, see top panel). Interestingly, \cite{Bresolin12} found that the extended disks of these galaxies (i. e. R$>$R$_{25}$) display an almost flat metallicity gradient. This result is consistent with the case of NGC 92, however, the ``outer disk'' (or tidal tail) of NGC 92 is considerably more metal rich (see Fig. \ref{MetalGradient}). 

\subsection{Searching for gas flows in the tail of NGC 92 using FP data}
\label{fabryperot_results}

One of the main mechanism discussed in the literature to explain flat metallicity gradients in the central parts of interacting galaxies is the mixture of gas given the gas inflow to the  central  regions of these systems \citep{Kewley06,Chien07,Kewley10,Rupke10,Werk11}. Interestingly, \cite{Rampazzo05} has successfully detected flows of warm gas between some interacting pairs of galaxies by using Fabry-Perot data cubes. 

A similar scenario can also explain the high oxygen abundances detected in tidal tails \citep{Rupke10}, given that at the same time that most of the gas goes inwards, there is some small fraction of the enriched gas from the center of the interacting system that is expelled to the outer parts of the galaxy, enhancing its metallicity. In this sense, NGC 92 becomes an ideal target to search for gas flows given that it is a clear case where an enhanced metallicity is measured all along the tail.  If ongoing gas mixing could be invoked as a cause for the enhanced metallicity measured for the tail of NGC 92 we would  then expect to find a velocity gradient in the tail of NGC 92 that would confirm the presence of a gas flow.   

We have used the H$\alpha$ Fabry-Perot data cube of  NGC 92, which was published by \cite{TorresFlores09} and has a spectral resolution of \mbox{$\sim$12 km s$^{-1}$}. In Fig. \ref{fabryperot} we plot the \textit{r\pr}-band image of the tidal tail of NGC 92, where we have included the H$\alpha$ profiles along this tail. In each case, the profiles were integrated over regions that are defined by red boxes. On each profile, we have included the radial velocity derived from a single Gaussian fit. By inspecting this figure, we do not detect a flow of ionised gas along the tail, although the inclination of this object (51\dg, as determined in \S \ref{distances}) is favourable for measuring line-of-sight motions, if these were present. Note that there is no massive galaxy close to the end of the tidal tail of NGC 92. In the case of \cite{Rampazzo05}, they found velocity gradients (around \mbox{130 km s$^{-1}$}) for only two of the objects of their sample, but velocity gradients are almost absent in the tidal tails of two other systems (gradients lower than \mbox{30 km s$^{-1}$}). Interestingly, like in the case of NGC 92, the two objects showing a small velocity gradient on the tails do not present massive galaxy at the tip of the tail.

However, there is still the possibility that the tail is in the plane
of the galaxy disk and, given that it coincides with the position angle of the galaxy
major axis, we would not be able to measure the radial component of the 
velocities, i.e., for points located along the major axis, the line-of-sight 
velocity gives no information on radial motions. Therefore, in that case, we could not say that the lack of gradient
means lack of gas flows. Yet another possibility for the enrichment of the tail as a whole
as well as the observed lack of metallicity gradient in the tidal
is the in-situ processing and formation of new stars 
at similar rates, all along the tail. Numerical simulations are needed to fully understand possible scenarios. All the results listed above are discussed in \S \ref{gas_flow_discussion}.

\section{Discussion}

In the following, we discuss the main results found in this paper.

\subsection{Inner and outer metallicity gradients for NGC 92}
\label{flat}

\subsubsection{Inner metallicity gradient,  R$<$R$_{25}$ }

In \S \ref{oxygen_gradient} we show that the metallicity gradient of NGC 92, within R$<$R$_{25}$ can be adjusted to a linear fit of slope beta=-0.0436$\pm$0.0158, which is similar to what was found for other interacting galaxies studied by  \cite{Bresolin12}, e.g. NGC 1512, with $\beta$=-0.034 dex kpc$^{-1}$ and \cite{Kewley10}, with $\beta$=-0.021 dex kpc$^{-1}$, for the average slope of eight galaxies. These gradients are statistically flatter than the average slope found by \cite{Zaritsky94} for the inner parts of  spiral galaxies, of $\beta$=-0.07 dex kpc$^{-1}$. The observed difference between the gradients within R$<$R$_{25}$ between normal spirals and interacting galaxies is not huge, but statistically significant \citep[e. g.][]{Rupke10b}. In fact, \cite{Rupke10} performed N-body/smoothed-particle hydrodynamics numerical simulations of interacting galaxies specifically to determine how interactions could affect the metallicity gradient of these systems. They find that the change in nuclear metallicity of an interacting galaxy is correlated with the amount of gas inflow to the central region, suggesting that it is due to the dilution of the central metallicity, as lower metallicity gas flows inward from the outskirts of the system. These confirm the observational result that galaxies in interacting systems have flatter metallicity gradients than non-interacting spiral galaxies. Rupke et al.$'$s results are also consistent with the study of \cite{Kewley06}, who found that galaxies in interacting pairs present lower central metallicities than field galaxies.

In the case of the interacting galaxy NGC 92, we derived a single linear fit to all observed points and alternatively we obtained a linear fit for regions located at R$<$R$_{25}$ and other linear fit for regions located at R$>$R$_{25}$ (see Fig. \ref{MetalGradient}). This last exercise is quite limited, given that we have just six regions inside R$_{25}$, however, it is useful to determine at least some trends in NGC 92. When we take into account the linear fit applied to all regions, we found that NGC 92 displays a flat metallicity gradient. A possible break in this relation can be found at R$_{25}$ ($\sim$10 kpc), where source TF17 displays a lower oxygen abundance. On the other hand, when we applied two linear fits for NGC 92, we found that the inner part of NGC 92 (R$<$R$_{25}$) displays a steeper slope than the case of a single linear fit. Interestingly, in this case, the oxygen abundances along the tidal tail remain constant. In any of the two cases (one or two linear fits for NGC 92), we found that the tidal tail of NGC 92 displays high oxygen abundances when compared with its central regions. This fact is confirmed when we compare NGC 92 with the extended disk galaxies NGC 1512 and NGC 3621 (see Fig. \ref{MetalGradient}). In these cases, their extended disks display low abundances when compared with the regions located in the tail of NGC 92. This suggests that the mechanism producing flat metallicity gradients in extended disks could be different from the process that produces flat gradients in tidal tails. Given that NGC 1512 is an interacting galaxy \citep{Bresolin12}, the strength of the interaction should be taken into account when studying metallicity gradients of interacting objects. In view of the results listed above, we will explore the best scenario to explain the metallicity gradient shown in NGC 92.

\begin{figure}
\includegraphics[width=\columnwidth]{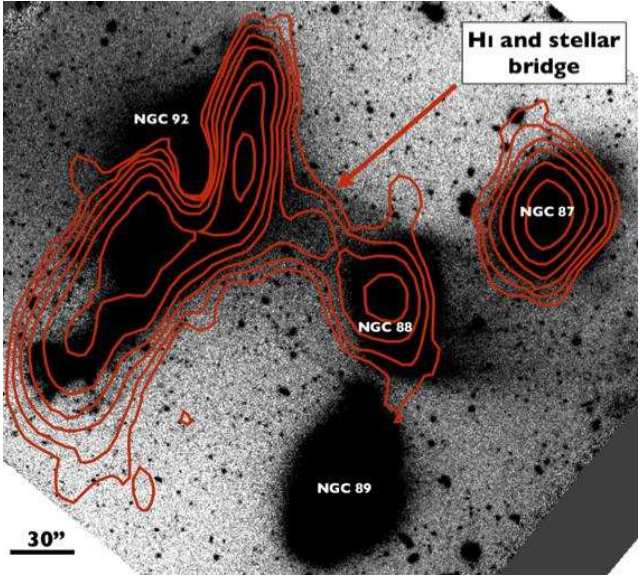}
\caption{\hi contours (taken from \citep{Pompei07}) overlaid on the Gemini \textit{r\pr}-band image of NGC 92 (image that is shown in high contrast in order to show some faint features). Contours represents surface masses of 5.4, 7.6, 10.8, 15.3, 21.6, 30.6 and 43.2 M$_{\odot}$ pc $^{-2}$ (see \S \ref{enhancement}). From this image it is possible to detect a stellar bridge between NGC 92 and its companion NGC 88, which is also visible in \hi gas \citep{Pompei07}.}
\label{hi}
\end{figure}

The flow of low-enriched gas into the central regions of merging galaxies can supposedly produce a burst of star formation in the nuclear regions and create flat metallicity gradients in the inner parts of interacting galaxies \citep{Kewley10}. In the case of NGC 92, \cite{Pompei07} found a strong central starburst, with a SFR=14.7 M$_{\odot}$ yr$^{-1}$ (as determined from ATCA radio continuum observations), which could be the source of the flatter gradient. In addition, \cite{Pompei07} found a gaseous bridge between NGC 92 and NGC 88, which is shown in Fig. \ref{hi}. Here, countours are overplotted on the Gemini/GMOS \textit{r\pr}-band image of the compact group that include the galaxy NGC 92. This image clearly shows a stellar bridge between NGC 92 and NGC 88. Given that there is no ionized gas along the bridge, we are not able to search for gas flows along these structures. The detection of a gaseous and stellar bridge between NGC 92 and NGC 88 suggests that both galaxies have experienced interaction in the past which could possibly have contributed to a central gas flow towards NGC 92 and a resulting starburst in its center. In fact, Gemini/GMOS data reveals that NGC 88 displays an oxygen abundance of \mbox{12+log(O/H)=8.54} (Scarano et al., in preparation). Therefore, gas flows from NGC 88 into the central region of NGC 92 could be producing a dilution of the original abundances of the nuclear region of this galaxy (as suggested by \citealt{Kewley06} and \citealt{Perez11}). This idea is supported by the high nuclear SFR displayed by NGC 92. 

\subsubsection{Is the metallicity high in the tail of NGC 92 due to star formation?}
\label{enhancement}

The tidal tail of NGC 92 displays several young star forming complexes. Given that this tail has an age of $\sim$190 Myrs \citep{Presotto10} we can expect that previous generations of \hii regions have been formed along this tidal tail. These previous generations of \hii regions could have enriched the interstellar medium of this tail with metals. A similar scenario was suggested as one of the possible enrichment mechanisms for the HI tail in the merger remnant NGC 2782 \citep{TorresFlores12}.  As shown in Fig. \ref{MetalGradient}, the oxygen abundance of the tidal tail (R$>$R$_{25}$) displays a gap with respect to the gradient shown by regions located inside R$_{25}$, i. e. the abundances in the tail are larger by $\sim$0.1 dex with respect to the fit of the inner region of NGC 92. Given that we already known this difference in oxygen abundance, together with the age of the tail, we are able to estimate the value for the yield necessary to increase the oxygen abundances in 0.1 dex. To do that, we have used equation 1 from \cite{Bresolin12}:

\begin{equation}
\frac{O}{H}=\frac{y_{0} \times t \times \Sigma_{SFR}}{\mu \times \Sigma_{HI}}
\end{equation}

where \textit{y$_{0}$} is the oxygen yield by mass, \textit{$\mu$}=11.81 is the conversion factor from number ratio to mass fraction and \textit{t} is the timescale for the star formation activity (which corresponds to 190 Myr, which is the age of the tail). In this case, we have estimated the $\Sigma_{SFR}$ from FUV/\textit{GALEX} data, where the flux along the tidal tails was measured with the task POLYPHOT in IRAF. The SFR was estimated by using the recipes given in \cite{IglesiasParamo06}. In the case of $\Sigma_{HI}$, we have converted the \hi flux given in \cite{Pompei07} in column density using the formulae given in \cite{Rohlfs00} and then we converted these values to units of M$_{\odot}$ pc$^{-2}$ (these values are listed in Fig. \ref{hi}). For the tail, we found $\Sigma_{SFR}$=8$\times$10$^{-5}$ M$_{\odot}$ yr$^{-1}$ kpc$^{-2}$. In general, this structure is located inside \hi regions of $\Sigma_{HI}$=10.8 M$_{\odot}$ pc$^{-2}$ (\mbox{N(HI)~$\sim1.3\times10^{21}$ cm$^{-2}$}), which is over the traditional threshold level for star formation (2-8 M$_{\odot}$ pc$^{-2}$ as determined by \citealt{Schaye04} and \mbox{N(HI)~$\sim4\times10^{20}$ cm$^{-2}$} as suggested by \citealt{Maybhate07} and \citealt{Mullan13}).

The results of this simple exercise are the following: We have found a value of \textit{y}$\sim$0.7, which is several times higher than the value suggested by \cite{Maeder92}, \textit{y$_{0}$}~$\sim$~0.01, if the SFR is derived using only the UV photometry. In order to have a complete estimation of the SFR, we should include the infrared emission (for example, at 24 $\mu$ m), however, no \textit{Spitzer} data were available for this target. We assume a SFR that is double that measured from the UV in the tail of NGC 92 (which is uncertain, given that the SFR calibrators are based on disk galaxies), and in that case the yield \textit{y}$\sim$0.3. Such high yields are expected to be found, given that it depends on the gas fraction and the latter is quite high for NGC 92 (note that $\textit{y}=Z/ln(\mu^{-1})$). If the yields values listed above are real, the star formation in the tail of NGC 92 has elevated the oxygen abundances along this structure. Otherwise, the results listed above suggest that the high oxygen abundances detected in the tidal tails of NGC 92 are not produced by a constant star formation along the age of the tail (assuming that the gas of the tail was expelled from the outer disk of NGC 92, i .e. radius$\sim$R$_{25}$).

\subsubsection{Origin of the high metallicities for R$>$R$_{25}$: gas flow and/or star formation }
\label{gas_flow_discussion}

\cite{Rupke10} found that a small fraction of enriched gas in interacting systems is ejected from the center of the galaxies outwards and this raises the gas metallicity all along tails. Indeed we find a very high metallicity in the tail region of NGC 92, which is close to solar, suggesting that Rupke et al.'s proposed scenario could be at work. However, if gas is moving from the galaxy center to the outskirts, one would expect to detect a velocity gradient along the tail, which is not observed (see \S \ref{fabryperot_results}).
 
If gas flow is not an ongoing process in the tail of  NGC 92,  an alternative or additional scenario that could explain the high metallicities found in the tail of NGC 92 could be continuous star formation through the evolution of several generations of stars.  As it was pointed out by \cite{Bresolin12}, the large gas reservoirs of some interacting systems could lead to high values of effective yield. As a simple exercise, we roughly calculated in the last section the value of the yield for NGC 92. We find a value of 0.3, under some assumptions, which although uncommonly high, may point to a  scenario of  enrichment due to the occurrence of star formation since the time of the tail formation. Either if this mechanism happens often or not, we do not know, but it is interesting that in this system, where we have high resolution kinematics information and detailed metallicity gradients, the commonly used explanation of ongoing gas flow as responsible for metallicity enhancement of the tail may not fully work (or may not be the only mechanism). 

In summary, we suggest that, in the case of NGC 92, the high oxygen abundances detected in this structure could be the result of enriched gas once expelled from the central region of NGC 92 which then remained in the tail forming stars at a very low level (as currently detected). However, the large amount of gas detected along this structure may hint at a large value for the yield (\textit{y}). Star formation processes may thus be responsible for enhancing the high oxygen abundances in a system in which enriched gas was expelled from the center to the tail at the time of the tail formation, 190 million years ago.

In addition, we find a flat metallicity gradient for the outskirts of NGC 92, which is possibly primordial, given that either the gas flow mechanism or the star formation scenario described above would have the function of  merely  enhancing the metallicity of the tail as a whole, keeping a flat gradient throughout the tail evolution.

\section{Summary and conclusions}

In this paper, we have studied the physical properties of 20 star-forming regions in the interacting galaxy NGC 92. Most of these objects were located in the optical/gaseous tidal tail of this galaxy. We found that these regions display young ages, which suggests that they were formed \textit{in situ}. Using new Gemini/GMOS spectroscopic data, we estimated the oxygen abundances for each region, and we study the metallicity gradient for this interacting galaxy. We found that NGC 92 presents a flat metallicity gradient with values close to solar. The usual scenario for explaining the high measurements of metallicities (close to solar) in tidal tails of merging galaxies is that of expulsion of enriched gas from the center outwards, throwing metals from the galaxies to their tail. Using H$\alpha$ Fabry-Perot data cubes, we did not detect a velocity gradient, and hence, no flow of ionized gas along the tidal tail of NGC 92, suggesting that this is not an {\it ongoing} process. However, there is also the possibility that the tail is in the plane 
of the disk and coincides with the major axis position angle of the galaxy, and in that case radial motions could
not be measured, if present. Gas expulsion possibly happened in the beginning of the interaction that formed the optical and HI tidal tail, which then started forming stars. Finally, probably star formation has been relevant in the chemical enrichment of this tidal tail since then.
We find a flat metallicity gradient for the outskirts of NGC 92, which is possibly primordial, i.e., the tidal tail had its metallicity enhanced as a whole, keeping a flat gradient, through its evolution.

\section{Acknowledgements}
We thank the anonymous referee for the useful comments that greatly improved this paper.
ST-F acknowledges the financial support of the Chilean agency
FONDECYT through a project ``Iniciaci\'on en la Investigaci\'on'', under contract
11121505 and ST-F also acknowledges the support of the project CONICYT PAI/ACADEMIA 7912010004. SSJ acknowledges FAPESP (Brazil) for the postdoc
grant 09/05181-8. HP thanks CNPq/CAPES for its financial support using the PROCAD project 552236/2011-0. This research has
made use of the NASA/IPAC Extragalactic Database (NED) which
is operated by the Jet Propulsion Laboratory, California Institute
of Technology, under contract with the National Aeronautics and
Space Administration. \textit{GALEX} is a NASA Small Explorer, launched in 2003 April. We
gratefully acknowledge NASA's support for construction, operation
and science analysis for the GALEX mission, developed in cooperation
with the Centre National d’ Etudes Spatiales of France and
the Korean Ministry of Science and Technology.

\bibliographystyle{mn2e}
\bibliography{main.bib}

\end{document}